\documentclass[aps,floatfix,twocolumn,nopacs,10pt,nofootinbib]{revtex4-1}
\usepackage[usenames,dvipsnames]{color}
\definecolor{dgreen}{rgb}{0.0, 0.5, 0.0}
\newcommand{\be}{\begin{equation}}
\newcommand{\ee}{\end{equation}}
\newcommand{\Iz}{\langle I_n \rangle}
\newcommand{\dIz}{\langle \delta I \rangle}
\newcommand{\dIzn}{\langle \delta I_n \rangle}
\newcommand{\dIzx}[1]{\langle \delta I (#1) \rangle}

\usepackage{graphicx}
\usepackage{amsfonts,amssymb,amsmath}
\usepackage{ulem}

\usepackage{hyperref}

\begin{document}


\title{
Dipole-like dynamical nuclear spin polarization around a quantum point contact}



\author{Peter Stano$^{1,2,3}$, Tomosuke Aono$^4$, Minoru Kawamura$^1$}
\affiliation{
$^1$RIKEN Center for Emergent Matter Science, 2-1 Hirosawa, Wako, Saitama, 351-0198 Japan\\
$^2$Department of Applied Physics, School of Engineering, University of Tokyo, 7-3-1 Hongo, Bunkyo-ku, Tokyo 113-8656, Japan\\
$^3$RCQI, Institute of Physics, Slovak Academy of Sciences, Dubravska cesta 9, 845 11 Bratislava, Slovakia,\\
$^4$Department of Electrical and Electronic Engineering, Ibaraki University, Hitachi 316-8511, Japan
}


\date{\today}

\begin{abstract}
We theoretically investigate the dynamical nuclear spin polarization in a quantum point contact (QPC) at finite magnetic field. We find that when the QPC is tuned to be spin selective, at the conductance of $e^2/h$, a finite bias induces a dipole-like (spatially anti-symmetric) nuclear polarization: at the QPC center the polarization is zero, while, for GaAs parameters, the nuclear spins down (up) are induced on the source (drain) side. We predict that the dipole-like  polarization pattern can be distinguished from a uniform polarization due to a qualitatively different response of the QPC conductance to the NMR field.
\end{abstract}


\maketitle

\section{Introduction}

There is a range of probe techniques based on polarized nuclear spins inside a solid-state material \cite{slichter, abragam}. In mesoscopic and nanoscopic physics related to spin, the advantage of using effective magnetic field of polarized nuclei is that it couples solely to electrons spins, without exerting orbital effects, offering a unique local and non-invasive probe \cite{ren2010:PRB}. However, unlike electrons which can be efficiently polarized by magnetic field, polarized light, ferromagnets, or spin-orbit interactions, it is not straightforward to polarize nuclei. The standard way is to transfer the polarization from electrons, known as the dynamical nuclear spin polarization (DNSP). It is done by using either non-equilibrium or equilibrium electronic spin polarization. In the first case, nuclei polarize as a reservoir into which the non-equilibrium electronic spin is dissipated \cite{overhauser1953:PR,lampel1968:PRL,aronov1976:JETPL,johnson2000:APL}. In the second case, the mutual electron-nuclear spin flip-flop transition is taken out of thermal equilibrium, by a resonant field \cite{abragam1978:RPP} or electrical bias voltage \cite{feher1959:PRL, komori2007:APL,kawamura2007:APL}.

We consider here a prototypical mesoscopic DNSP experiment, where a quantum point contact (QPC) exhibits spin-selective transport. This arises if a large---several Tesla---magnetic field is applied, and the QPC is gated to around 
1/2 of conductance quantum $2e^2/h$. We investigate what nuclear polarization is expected in this case and how it can be detected. For the latter, we consider the QPC conductance changes upon applying NMR, so-called resistively detected NMR (RD-NMR) \cite{dobers1988:PRL}. Variations of this setup have been already considered in experiments and theory \cite{wald1994:PRL, cooper2008:PRB, ren2010:PRB, kawamura2015:PRL, singha2017:PRB, fauzi2017:PRB}, focusing on various phenomena such as conductance hysteresis, nuclear-spin relaxation, the DNSP mechanism, or ``dispersive'' RD-NMR line shapes. We, however, feel that despite all these investigations, certain general aspects remained obscure. Their clarification is our main goal.

Namely, we deem the considered setup as a generic picture of the DNSP arising at a spin-sensitive electron scatterer. We point out that there is a fundamental difference between a scatterer with leads which themselves do, or do not, contain electronic {\it nonequilibrium} spin polarization. In the first case, this nonequilibrium polarization flows from electrons to nuclei, and the scatterer plays a minor role. In the second case, which is perhaps more often relevant in experiments, and is also of prime interest for us, the result is different and can be summarized as the following. An electrically biased spin-selective scatterer disturbs the electronic spin density, bringing it locally out of equilibrium, into a spatially non-uniform pattern. The nuclei polarize locally according to this local non-equilibrium electronic spin density. Since, however, the scatterer only redistributes the electronic spin, but does not create or dissipate it, the total amount of produced nuclear polarization is zero. 

This general behavior is exemplified by a spin-selective QPC, which we study in detail below. We find that, proviso certain mild and realistic conditions, it develops a spatially antisymmetric DNSP pattern, with zero nuclear polarization at the QPC top. We call this pattern a dipole-like nuclear polarization. Only with additional rather strong asymmetry sources, either in the structure geometry or, as we identify here, in the nuclear relaxation, the polarization on one side might dominate, resulting in an overall net nuclear polarization. However, we do not deem such asymmetries to be typically the case, and expect a dipole-like DNSP around a QPC at low temperatures. This constitutes our main result. We also note that it is a spin analogy of the charge resistivity dipole of Ref.~\onlinecite{landauer1975:ZPB}. The theory of DNSP, and the elucidation of the conditions for the polarization spatial symmetry, is the content of Sec.~II. 

The second rather complicated issue has to do with the detection of the established nuclear polarization. With its volume being too small to be seen directly as a standard NMR signal, we consider the QPC itself as a probe \cite{kawamura2015:PRL}. The detection proceeds by monitoring the QPC conductance upon scanning the microwave frequency. At a resonance, the NMR field depolarizes nuclear spins, which decreases the Overhauser field contribution to the total Zeeman energy. If the QPC conductance is sensitive to it, the presence of nuclear polarization can be detected, forming basis for RD-NMR methods. To infer something about the nuclear polarization beyond its presence is, however, highly non-trivial. Indeed, the response of the conductance to a change in the Overhauser field can have both signs, depending on the applied bias, gate voltage, external field, and electron-electron interactions \cite{stern2004:PRB,ren2010:PRB,fauzi2017:PRB}. It is clearly even more so, if the Overhauser field is spatially inhomogeneous, which is the case here.

We derive the QPC conductance response analytically for a simple model of non-interacting electrons at zero temperature and zero bias in Sec.~III. We find that the conductance response differs qualitatively for a uniform and a dipole-like nuclear polarization, if observed both below and above the 1/2 conductance. We then compare this prediction with the experiment of Ref.~\onlinecite{kawamura2015:PRL} and find a sign discrepancy. By modeling the detection of a nuclear dipole numerically, we find that the discrepancy can be resolved by including strong electron-electron interactions. We then arrive at a qualitative, but not quantitative, agreement with that experiment in Sec.~IV. Finally, we note that the presence of the nuclear dipole could be revealed as a "dispersive lineshape" of the conductance response if probed as a function of the NMR frequency. This is discussed in Sec.~V.

To smoothen the text flow, we delegate several auxiliary results to appendices. In App.~A we present our numerical model. In App.~B we demonstrate that a sign reversal due to Coulomb interactions is robust. In App.~C we derive the electron-nuclear spin flip-flop rates, Eqs.~\eqref{eq:Iz1}. In App.~D we derive the microscopic equilibrium relation, Eq.~\eqref{eq:eq}. Finally, in App.~E we estimate the asymmetry of the polarization around the QPC due to difference in electron velocities, and leads' cross sections.

\section{DNSP dipole}

\label{sec:dipole}

In this section, we construct the theory of the DNSP for a one-dimensional spin-selective scatterer.

\subsection{Model of the coupled electron-nuclear system}

\begin{figure}
\includegraphics[width=\columnwidth]{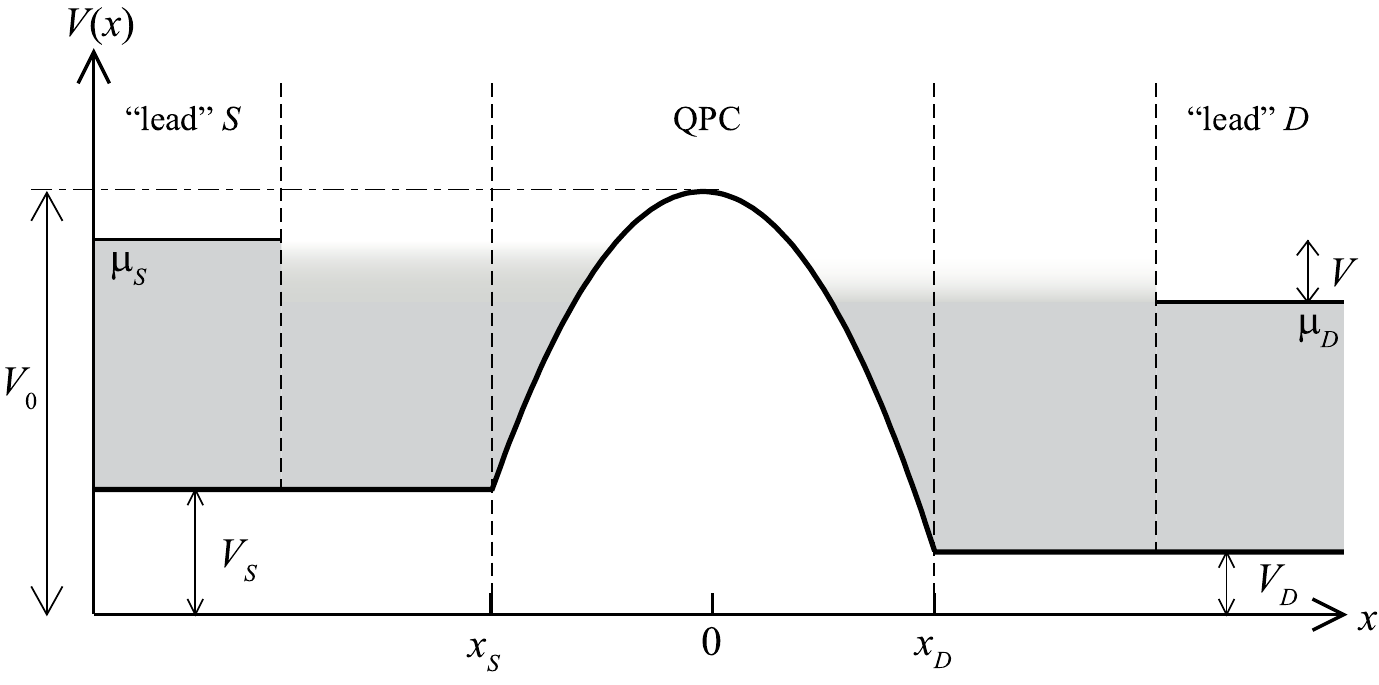}
\caption{\label{fig:scatterer}
Illustration of the setup we consider to model transport through a QPC. The thick line shows the potential $V(x)$ as a function of position. To the left of $x_S$, and the right of $x_D$, the potential is constant, defining the left and right lead, respectively. The part between these two points represents the scatterer. The shaded areas denote the states occupation in the two leads, up to the respective chemical potentials, which differ by the applied bias. For illustration, by a sharp, and shaded occupations around the chemical potential we distinguish, respectively, the equilibrium and non-equilibrium electronic distributions (see text for explanation). The non-equilibrium distributions of electron spins extend from the QPC over the distance given roughly by the spin diffusion length.
}
\end{figure}
We consider dynamical nuclear spin polarization arising in a biased QPC in the presence of a finite magnetic field. To this end, we consider electrons moving in a one-dimensional potential,
\be
V(x) = \left\{
\begin{tabular}{ll}
$V_S$&if $x<x_S$,\\
$V_{Q}(x)$&if $x_S<x<x_D$,\\
$V_D$&if $x_D<x$,\\
\end{tabular}
\right.
\label{eq:V}
\ee
separating the space into the source lead, scatterer, and the drain lead, respectively.\footnote{Rather than by a flat potential, the leads are defined as regions where electronic states are in equilibrium, with occupations given by the Fermi functions [see Eq.~\eqref{eq:n} below]. For non-interacting electrons, which are not backscattered once they reach the flat region, the scatterer can be shrank to the space between $x_S$ and $x_D$. For an interacting model, one should include also enough space for the equilibration.} While the potential of the scatterer is in general unknown, the observed conductance quantization of a typical well-behaved QPC \cite{patel1991:PRB} can be described by the quadratic model \cite{buttiker1990:PRB},
\be
V_Q(x) = V_0 - \frac{1}{2}m \omega^2 x^2,
\label{eq:VQPC}
\ee
with $V_0$ the energy at the potential maximum (the center of the QPC) located at $x=0$, $m$ the electron effective mass, and $\hbar \omega$ an energy parametrizing the potential curvature, alternatively expressed using length $l_{\rm QPC} = \surd{\hbar /m \omega}$. The QPC conductance is tunable through the gate voltage $V_g$, which shifts the top of the barrier, $\delta V_0 = \alpha e \delta V_g$. Here, $\alpha$ is the lever arm converting the gate voltage to the potential energy, typically $\alpha \sim 0.01$--$0.1$. The considered geometry is depicted in Fig.~\ref{fig:scatterer}.

The electron scattering is spin sensitive due to the Zeeman energy $\epsilon_z = |g \mu_B B|$ resulting from the applied magnetic field ${\bf B}$, in proportion to the Bohr magneton $\mu_B$ and the electron $g$ factor $g$, negative in GaAs. We neglect the effects of the spin-orbit interaction as, first, intrinsically weak in GaAs, and, second, its effects being further suppressed in a one-dimensional geometry in both tunneling \cite{stano2010:PRB}, and quasi-ballistic regimes \cite{stano2011:PRL}.\footnote{Combined effects of the spin-orbit interaction and nuclear polarization were considered in Refs.~\onlinecite{tripathi2008:EPL,nesteroff2004:PRL}.} We thus arrive at
\be
H_0 = \frac{p^2}{2m} + V(x) + \frac{g \mu_B}{2} \boldsymbol{\sigma} \cdot {\bf B},
\label{eq:H0}
\ee
as the unperturbed Hamiltonian for electrons.

We now define the scattering states as the eigenstates of $H_0$ with the boundary condition being an incoming wave of unit amplitude in one of the leads. We denote
\be
\Psi_{\epsilon l \sigma}(x) = \langle x| \Psi_{\epsilon l\sigma}\rangle,
\label{eq:state}
\ee
as the wave-function amplitude of the scattering state with energy $\epsilon$, originating in the lead $l \in \{ S, D \}$, with spin $\sigma \in \{ \uparrow, \downarrow \} \equiv \{ +1, -1\}$, corresponding, respectively, to spinors parallel and antiparallel to ${\bf B}$. Below, we use $\overline{l}$ to denote the lead opposite to $l$, and similarly for the spin. The amplitudes of a scattering state in the leads are described by the scattering matrix $S_{l^\prime l}^{\sigma^\prime \sigma}(\epsilon)$, relating the outgoing current amplitude in lead $l^\prime$ with spin $\sigma^\prime$ to the incoming current amplitude in lead $l$ for electrons with spin $\sigma$ at energy $\epsilon$ \cite{gasparian1996:PRA}. 
The system definition is completed by specifying the scattering states occupations
\be
n_{l\sigma}(\epsilon)= \left\{ \exp[\beta(\epsilon - \mu_{l\sigma})] +1 \right\}^{-1},
\label{eq:n}
\ee
which depend on the temperature $T$,  parameterized by energy $\beta^{-1} = k_B T$, with $k_B$ the Boltzmann constant, and the leads' chemical potentials $\mu_{l\sigma}$. Here, we kept the possibility of different chemical potentials for the two spin species, representing a lead with non-equilibrium spin accumulation \cite{meair2011:PRB}. However, we will consider this option only in Sec.~\ref{sec:non-eq}. Everywhere else we take them spin independent, and
for a notational convenience, symmetrically displaced from the Fermi energy by half of the applied bias voltage $V$, $\mu_{S/D} = \mu_{\rm F} \pm eV/2$.

The conservation of the spin along the magnetic field yields the scattering matrix diagonal in the spin index, $S_{l^\prime l}^{\sigma^\prime \sigma}(\epsilon) \propto \delta_{\sigma^\prime \sigma}$. In addition, the spin dependence enters only through the Zeeman energy, by which the kinetic energy differs for the two spin species at the same total energy $\epsilon$:
\be
\epsilon_{\rm kin}^{\sigma}(\epsilon) = \epsilon - \frac{\sigma}{2} \epsilon_z.
\ee 
Then, the scattering state amplitudes in the leads are
\begin{subequations}
\begin{eqnarray}
\Psi_{\epsilon S \sigma}(x)  &=& \left\{
\begin{tabular}{ll}
$e^{i k_{S\sigma} x} + S_{SS}(\epsilon_{\rm kin}^{\sigma}) e^{- i k_{S\sigma} x}$,&if $x < x_S$,\\
$\sqrt{\frac{v_{S\sigma}}{v_{D\sigma}}}S_{DS}(\epsilon_{\rm kin}^{\sigma}) e^{i k_{D\sigma} x} $,&if $x > x_D$,\\
\end{tabular}
\right.\\
\Psi_{\epsilon D \sigma}(x)  &=& \left\{
\begin{tabular}{ll}
$e^{-i k_{D\sigma} x} + S_{DD}(\epsilon_{\rm kin}^{\sigma}) e^{i k_{D\sigma} x}$,&if $x > x_D$,\\
$\sqrt{\frac{v_{D\sigma}}{v_{S\sigma}}}S_{SD}(\epsilon_{\rm kin}^{\sigma}) e^{-i k_{S\sigma} x} $,&if $x < x_S$,\\
\end{tabular}
\right.
\end{eqnarray}
\label{eq:ss}
\end{subequations}
where we introduced the wave vectors $k$ and velocities $v$, both spin and possibly lead dependent, by
\be
\frac{\hbar^2 k_{l\sigma}^2}{2m} \equiv \frac{1}{2} m v_{l\sigma}^2=\epsilon - \frac{\sigma}{2}\epsilon_z -V_l,
\label{eq:kv}
\ee
and we will also use the density of states $g_{l\sigma}=1/2\pi\hbar v_{l\sigma}$. 

For spin-preserving and single-subband scattering, all four elements of the scattering matrix $S$ are given by a single parameter, the transition probability $T$ \cite{buttiker1985:PRB,buttiker1986:PRL},
\be
|S_{\overline{l}l}|^2 = T, \qquad |S_{l l}|^2 = R=1-T.
\label{eq:Ssym}
\ee
This is a key property for the discussion below and we stress that both the above conditions are necessary. Namely, were the scattering spin dependent, the elements of the scattering matrix would be matrices \cite{bardarson2007:PRL}, without any symmetry relations in general.\footnote{Namely, in this case, such symmetry relations for the scattering matrix would require additional symmetry in the Hamiltonian, for example, the time-reversal symmetry \cite{zhai2005:PRL, kiselev2005:PRB}.} Were the leads not single subband, the unitarity alone is not enough to reduce the scattering matrix to a single parameter such as in Eq.~\eqref{eq:Ssym}. We note that for the model given in Eq.~\eqref{eq:VQPC}, the transition probability can be calculated analytically \cite{buttiker1990:PRB}:
\be
T(\epsilon_{\rm kin}^\sigma)=\left\{1+\exp\left[-2\pi \frac{\epsilon_{\rm kin}^\sigma-V_0}{\hbar \omega}  \right]\right\}^{-1}.
\label{eq:TQPC}
\ee
However, below we do not rely on a specific form of $T$. We will only use that the transition probability has a form similar to that given in the previous equation: it is a monotonically increasing function of the kinetic energy, which at zero bias and temperature changes from 0 to 1 over an energy of the order of $\hbar \omega$. The latter therefore represents the energy resolution of the QPC.

We are interested in the effects of the electrical current on nuclear spins. The electrons and nuclei are coupled by the Fermi contact interaction,
\be
H_I =  v_0 \sum_n A_n \delta({\bf r} - {\bf r}_n) \boldsymbol{\sigma} \cdot {\bf I}_n.
\label{eq:HI}
\ee
Here, the discrete index $n$ labels nuclear spins with the corresponding spin operators ${\bf I}_n$, located at positions ${\bf r}_n$, $A_n$ is the material constant which depends on the nuclear isotope, and $v_0$ is the volume per nuclear spin, in a zinc-blende material equal to $a_0^3/8$, with the lattice constant $a_0$. In Eq.~\eqref{eq:HI}, the positions of both nuclei and electrons are three-dimensional vectors. In connection to this, one needs to consider the transverse wave-function components of the electronic states. Assuming for simplicity that the three-dimensional wave function can be factorized\footnote{To adopt this form, we are motivated by the following consideration: We assume that the wave function can be approximated as separable $\Phi(x,y,z) \approx \Psi(x) \times \phi(y,z;l_y, l_z)$, where $l_y,$ and $l_z$ are parameters of the transverse profile (e.g., confinement lengths). They are assumed to be weakly dependent on $x$, so that as the electron moves along $x$, the changes of these parameters are followed adiabatically. The $x$ dependence of these parameters results in the $x$ dependence of the transverse profile $\phi$.} $\Phi(x,y,z) \approx \Psi(x) \times \phi(x,y,z)$, we define a normalized transverse density $\rho({\bf r}) \equiv | \phi({\bf r})|^2$:
\be
\int {\rm d}y {\rm d}z \rho(x,y,z)=1,
\label{eq:densitynormalization}
\ee
and, using $\rho_n \equiv \rho({\bf r}_n)$, write
\be
H_I = v_0 \sum_n A_n \rho_n \delta(x - x_n) \boldsymbol{\sigma} \cdot {\bf I}_n.
\ee
We separate the effects of the transverse coordinates by defining a dimensionless quantity
\be
t_n = {\rho_n} {S_\perp} (x=0),
\label{eq:tn}
\ee
where a cross section at the longitudinal coordinate $x$,
\be
S_\perp(x) =  \frac{1}{\int {\rm d}y {\rm d}z | \phi(x,y,z)|^4},
\ee
loosely defines an area within which the electron interacts with nuclei appreciably \cite{merkulov2002:PRB}.
With this
\be
H_I = \frac{A v_0}{S_\perp} \sum_n t_n \delta(x - x_n) \boldsymbol{\sigma} \cdot {\bf I}_n, 
\label{eq:HIfinal}
\ee
where $S_\perp \equiv S_\perp(x=0)$, and we assumed a homonuclear system, $A_n=A$, for simplicity (although a possible variation of these factors could be accounted for in $t_n$). Equation \eqref{eq:HIfinal} is the form for the hyperfine interaction which we use in what follows.

Unless stated otherwise, we use the parameters of bulk GaAs, that is, $a_0=0.565$ nm, $g=-0.44$, $m=0.067$ $m_e$, with $m_e$ the free-electron mass, $A = 45$ $\mu$eV, as an average over natural isotopes, and $I=3/2$ for the nuclear spin magnitude. We further take $\hbar \omega \sim 0.5$--$1.0$ meV, corresponding to $ l_{\rm QPC} \sim 30$--$50$ nm, and $S_\perp \approx 200$ nm$^2$ as representative values for a typical QPC.

\subsection{Electron-assisted nuclear dynamics}

With the above definitions, we derive the master equation for the $n$-th nuclear spin in the lowest order in the hyperfine coupling $A$, by treating the hyperfine interaction as a perturbation causing transitions between the unperturbed scattering states. We get 
\be \begin{split}
\partial_t \Iz = \frac{2}{3}I(I+1) &\Big( W_{\downarrow\uparrow}({\bf r}_n) - W_{\uparrow\downarrow}({\bf r}_n) \Big)\\
-\Iz  &\Big( W_{\downarrow\uparrow}({\bf r}_n) + W_{\uparrow\downarrow}({\bf r}_n) \Big),
\end{split}
\label{eq:Iz1}
\ee 
where we assumed a small polarization $\Iz \ll I$,\footnote{The equation is also exact for $I=1/2$ and an arbitrary polarization. For the general case of an arbitrary nuclear spin and polarization, the nuclear-spin-magnitude-dependent prefactors on the right-hand side, such as the given value $2I(I+1)/3$, depend on the density matrix of the nuclear spin, so that additional assumptions on its form are necessary to evaluate these factors. See App.~\ref{app:rates} for details. These complications are of no relevance to us, as the polarizations are indeed small, and, in addition, we are interested in qualitative properties of the nuclear polarization, most importantly, its sign and spatial profile, rather than in its quantitative value.} 
defined by $\Iz \equiv \langle {\bf I}_n \cdot {\bf B}/B\rangle$, that is as the projection of the spin polarization on the direction of ${\bf B}$. Finally, the angular brackets denote the average over the system density matrix.
Equation \eqref{eq:Iz1} gives two contributions to the nuclear dynamics, the polarization, and decay, as the first and second terms, respectively. 
The rates in Eq.~\eqref{eq:Iz1} are defined by
\be
\begin{split}
&W_{\overline{\sigma}\sigma}  ({\bf r}_n) =     \sum_{l l^\prime} \int \frac{{\rm d}\epsilon}{\hbar} W_0 (\epsilon) t_n^2
\\&  \qquad \times |\Psi_{\epsilon l \sigma}(x_n)|^2|\Psi_{\epsilon^\prime l^\prime \overline{\sigma}}(x_n)|^2  n_l(\epsilon)  [1 - n_{l^\prime}(\epsilon^\prime) ],
\end{split}
\label{eq:Wss}
\ee
the rate of the electron spin to flip from $\sigma$ to $\overline{\sigma}$ (corresponding to a nuclear spin change by $2\sigma$). Further, $\epsilon^\prime=\epsilon+\epsilon_z^N$, 
with $\epsilon_z^N=g_N \mu_N B$ being the nuclear Zeeman energy, defined by the nuclear $g$ factor $g_N$, and the nuclear magneton $\mu_N$, and 
\be
W_0 (\epsilon) = 2\pi \frac{(v_0 A)^2}{S_\perp^2} g_{l\sigma}(\epsilon)g_{l'\sigma'}(\epsilon').
\label{eq:W0}
\ee
is a dimensionless factor. Neglecting the spin, lead and bias dependence of the density of states,
\be
W_0 = \frac{1}{2\pi} \left( \frac{m}{2\mu_{\rm F}} \right)\left( \frac{A v_0}{\hbar S_\perp} \right)^2.
\ee
Finally, we also define renormalized rates
\be
w_{\overline{\sigma}\sigma} (x_n) = \frac{1}{W_0 t_n^2} W_{\overline{\sigma}\sigma}({\bf r}_n),
\label{eq:wss}
\ee 
which are stripped of the overall scale, and the transverse profile dependence, what makes them dependent only on the longitudinal coordinate.

We delegate the derivation of these equations to App.~\ref{app:rates}, as a straightforward generalization of the Slichter formula \cite{slichter,berg1990:PRL} to a spatially inhomogeneous system in a finite magnetic field \cite{singha2017:PRB}. Here, instead, we discuss the physical meaning of Eq.~\eqref{eq:Wss}: it gives the rate of nuclear spin change by 2$\sigma$ as the product of a probability to have an electronic occupied state with spin $\sigma$ times the probability of an unoccupied state with the opposite spin, both evaluated at the position of the particular nucleus ${\bf r}_n$. The only dimensionful expression in Eq.~\eqref{eq:Wss} is the integration measure, with units of a particle current. The strength of the process is parametrized by a dimensionless quantity $W_0$. Taking GaAs parameters, $W_0 \times \epsilon_0/\hbar$ evaluates to $9.2 \times 10^{-6}$ s$^{-1}$ for $\epsilon_0=k_B T$ at the temperature of 0.1 K, and to $3.2 \times 10^{-4}$ s$^{-1}$ for $\epsilon_0=e V$ at the bias voltage of $300$ $\mu$V. These two values give rough estimates for the nuclear spin equilibration and polarization rate due to electrons to be expected in the experiment of Ref.~\onlinecite{kawamura2015:PRL}.

\subsection{The polarization and equilibration rates}

For brevity, we omit the nuclear spin position argument $({\bf r}_n)$ from all rates in this subsection. We now split the rate in Eq.~\eqref{eq:Wss} to two terms, 
\be
W_{\overline{\sigma}\sigma}= W_{\overline{\sigma}\sigma}^{(1)} + W_{\overline{\sigma}\sigma}^{(2)}, 
\label{eq:W1W2}
\ee
according to whether the incoming leads of the initial and the final scattering state are the same, $l^\prime = l$, 
\be
\begin{split}
W_{\overline{\sigma}\sigma}^{(1)} & = W_0  t_n^2 \sum_{l} \int \frac{{\rm d}\epsilon}{\hbar}
\\&\times|\Psi_{\epsilon l \sigma}(x_n)|^2 |\Psi_{\epsilon^\prime l \overline{\sigma}}(x_n)|^2 n_l(\epsilon) [1 - n_{l}(\epsilon^\prime) ],
\end{split}
\label{eq:W1}
\ee
or opposite,  $l^\prime = \overline{l}$,
\be
\begin{split}
W_{\overline{\sigma}\sigma}^{(2)} &= W_0  t_n^2 \sum_{l} \int \frac{{\rm d}\epsilon}{\hbar}
\\&\times|\Psi_{\epsilon l \sigma}(x_n)|^2 |\Psi_{\epsilon^\prime \overline{l} \overline{\sigma}}(x_n)|^2 n_l(\epsilon) [1 - n_{\overline{l}}(\epsilon^\prime) ].
\end{split}
\label{eq:W2}
\ee
Using the identity 
\be
\frac{n(\epsilon)}{1-n(\epsilon)} = e^{-\beta(\epsilon-\mu)},
\label{eq:nsym}
\ee
we find that, irrespective of the applied bias,
\be
W_{\overline{\sigma}\sigma}^{(1)}  = W_{\sigma\overline{\sigma}}^{(1)}  \exp(-\sigma \beta \epsilon_z^N).
\label{eq:eq}
\ee
Since this is a condition of thermal equilibrium, we interpret $W^{(1)}$ as the equilibration rate and $W^{(2)}$ as the polarization rate. We note that such a distinction is only qualitative, since, for example, at zero bias $W^{(2)}$ also fulfills Eq.~\eqref{eq:eq} and therefore contributes only to the equilibration. 
We are motivated by considering the opposite regime, $eV \gtrsim k_B T, \epsilon_z^N$. In the limit $\epsilon_z^N \ll k_B T \ll eV$, the two rates are
\begin{subequations}
\begin{eqnarray}
w_{\overline{\sigma}\sigma}^{(1)} & = &  - k_B T \sum_{l} \int \frac{{\rm d}\epsilon}{\hbar}
|\Psi_{\epsilon l \sigma}|^2 |\Psi_{\epsilon l \overline{\sigma}}|^2 \partial_\epsilon n_l(\epsilon),
\label{eq:wlimit1}\\
w_{\overline{\sigma}\sigma}^{(2)} & = & \sum_{l} \int_{\mu_{\rm F}-eV/2}^{\mu_{\rm F}+eV/2} \frac{{\rm d}\epsilon}{\hbar}
|\Psi_{\epsilon l \sigma}|^2 |\Psi_{\epsilon \overline{l} \overline{\sigma}}|^2,
\label{eq:wlimit2}
\end{eqnarray}
\label{eq:wlimit}
\end{subequations}
so that $W^{(1)}\propto k_BT $ has the scaling of the Korringa law, while $W^{(2)}\propto e V $ scales with the applied voltage, the distinction expected for the equilibration, and polarization, respectively. In this regime, the polarization rate $W^{(2)}$ does not fulfill Eq.~\eqref{eq:eq}, and can become much larger than $W^{(1)}$, leading to a substantial non-equilibrium polarization (dynamical nuclear spin polarization). 

The possibility of $W^{(2)} \gg W^{(1)}$ necessitates to consider additional sources of equilibration for nuclei. Typical examples are the nuclear dipole-dipole diffusion, and spin -impurity and -lattice relaxations \cite{abrahams1957:PR, klauder1962:PR, lowe1967:PR, mcneil1976:PRB, paget1982:PRB, deng2005:PRB, gong2011:NJP, stano2013:PRB}. Assuming that these channels are not influenced by the applied voltage, the corresponding rates, which we denote as $W^{(0)}$, also fulfill the microscopic equilibrium condition, Eq.~\eqref{eq:eq}. We add such rates into Eq.~\eqref{eq:Iz1} and obtain the steady-state solution
\be
\Iz =  \frac{2}{3}I(I+1) \frac
{\sum_{i=0,1,2} \left( W^{(i)}_{\downarrow\uparrow} - W^{(i)}_{\uparrow\downarrow} \right)} 
{\sum_{i=0,1,2} \left( W^{(i)}_{\downarrow\uparrow} + W^{(i)}_{\uparrow\downarrow} \right)} ,
\label{eq:Iz3}
\ee 
where $W^{(0)}_{\downarrow \uparrow}$ should be understood as the nuclear spin-increasing rate, following loosely the notation introduced in Eq.~\eqref{eq:Wss}, and analogously for the spin-decreasing rate.
For zero bias, where all rates fulfill Eq.~\eqref{eq:eq}, the nuclear polarization is 
\be
\Iz_0 = \frac{2}{3}I(I+1) \frac{1-\exp(-\beta \epsilon_z^N)}{1+\exp(-\beta \epsilon_z^N)} \approx \frac{I(I+1)}{3}\frac{\epsilon_z^N}{k_B T},
\label{eq:Iz2}
\ee
the Curie's law of magnetization. 
However, interested in the non-equilibrium nuclear spin polarization at temperatures where the equilibrium one is small, we neglect from here on the nuclear Zeeman energy in all rates, setting $\epsilon^\prime = \epsilon$. The resulting nuclear polarization should be then understood as a departure from the thermal value
\be
\dIzn \equiv \Iz -\Iz_0 \approx  \frac{2}{3}I(I+1) \frac{W^{(2)}_{\downarrow\uparrow} - W^{(2)}_{\uparrow\downarrow} }
{\Gamma},
\label{eq:Iz4}
\ee 
where $\Gamma = \sum_{i,\sigma} W_{\sigma\overline{\sigma}}^{(i)}$, is the total equilibration rate, contributed by the hyperfine, and other interactions. We replace it by a phenomenological value, the nuclear spin equilibration rate measured experimentally.

We now reinstate the coordinate variable, and for further convenience, we define the polarization in the middle of the channel as
\be
\dIzx{x}=\dIzx{x,0,0}.
\ee
It allows us to write 
\be
\dIzn \equiv \dIzx{x} \frac{\rho_n^2}{\rho^2(x,0,0)},
\ee
relating the polarization along the transverse coordinates to the polarization in the center.

\subsection{The spatial symmetry of the polarization rates}

We now make the crucial observation regarding the symmetry of the rates in Eq.~\eqref{eq:W1W2}. From Eqs.~\eqref{eq:ss} and \eqref{eq:Ssym}, it follows that the scattering states {\it in the leads} fulfill
\be
|\Psi_{\epsilon l \sigma}(x)|^2 = |\Psi_{\epsilon \overline{l} \sigma}(-x)|^2,
\label{eq:Psisym}
\ee
if the velocities in two leads are, at a given energy, the same, what we assume,\footnote{The difference of the velocities is bounded by $|v_S-v_D| / |v_S+v_D| \leq eV /\mu_{\rm F}$, which is very small for typical parameters. For illustration, in App.~\ref{app:geometry asymmetry} we consider the asymmetries due to different electron velocities, and show that the expected effects are indeed negligible.} and we also drop the interference terms $\propto \exp(2 i k_{l\sigma} x)$.\footnote{These oscillatory interference terms do not contribute to the current \cite{buttiker1992:PRB}, but survive in the electron density, which is more important here. One can consider a spatial average of the rate, or polarization, over a distance larger than the electron Fermi wavelength, upon which the interference terms average to zero. Since typically the nuclear diffusion length is larger than the period of these oscillatory terms, considering the averaged nuclear polarization is actually more physical.} The most simple case to visualize the above equation is to consider the whole structure (including the scatterer) as inversion symmetric, $V(x) = V(-x)$. In this case, Eq.~\eqref{eq:Psisym} holds for any $x$. However, we stress that for our purposes we do not require such high symmetry, and allow for an asymmetric scatterer, in which case we only assume that both $x$ and $-x$ are within the opposite leads. Equation \eqref{eq:Psisym} follows from the unitarity of the scattering matrix \cite{buttiker1988:IBM}, and therefore holds also in presence of orbital effects of magnetic field.\footnote{The orbital effects of magnetic field can influence the scattering states inside the leads, for example, by shifting them by the Lorentz force \cite{ford1989:PRL} according to their velocity direction. This has no effect on the relation in Eq.~\eqref{eq:Psisym}, if these shifts are absorbed as the correspond changes of the transverse part of the scattering state, $\phi(x,y,z)$, defined above Eq.~\eqref{eq:densitynormalization}. }

Using Eq.~\eqref{eq:Psisym}, we obtain that the polarization rate in the leads fulfills
\be
w^{(2)}_{\sigma\overline{\sigma}}(x) = w^{(2)}_{\overline{\sigma}\sigma}(-x),
\label{eq:w2sym}
\ee
so that it is the same upon switching leads and inverting spins. There is no such symmetry relation for the equilibration rate. Indeed, from Eq.~\eqref{eq:wlimit1} we get a spatial asymmetry (here we assume $x>0$)
\begin{subequations}
\label{eq:w1sym}
\be
\begin{split}
w_{\overline{\sigma}\sigma}^{(1)}(x) &- w_{\overline{\sigma}\sigma}^{(1)}(-x) = - {2 k_B T}   \int \frac{{\rm d}\epsilon}{\hbar}  \\ & 
\times \left[ R(\epsilon_{\rm kin}^\sigma) + R(\epsilon_{\rm kin}^{\overline{\sigma}}) \right] \partial_\epsilon \left[n_D(\epsilon)-n_S(\epsilon) \right],
 \end{split}
\label{eq:w1sym1}
\ee
which is not zero in general. In fact, if the transition probability is a monotonically increasing function of energy, such as the one in Eq.~\eqref{eq:TQPC}, it is straightforward to see that the above asymmetry has the same sign as $\mu_S-\mu_D$, the bias voltage. In other words, the equilibration rate is always larger in the drain lead. For completeness, we also give
\be
\begin{split}
w_{\overline{\sigma}\sigma}^{(1)}(x) & + w_{\overline{\sigma}\sigma}^{(1)}(-x) = - {2 k_B T}   \int \frac{{\rm d}\epsilon}{\hbar} \\ & 
 \left[ 1 + R(\epsilon_{\rm kin}^\sigma) R(\epsilon_{\rm kin}^{\overline{\sigma}}) \right] \partial_\epsilon \left[n_D(\epsilon)+n_S(\epsilon) \right],
 \end{split}
\label{eq:w1sym2}
\ee
\end{subequations}
the symmetric part of the equilibration rate.

\begin{figure}
\includegraphics[width=\columnwidth]{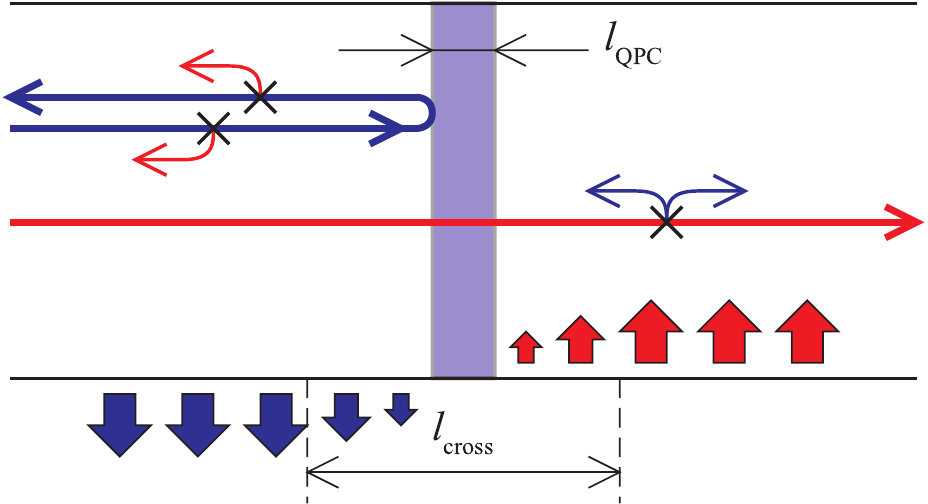}
\caption{\label{fig:dipole}
Illustration of the DNSP with a nuclear dipole-like profile. Electrons at an energy within the bias window are impinging from the left lead towards the scatterer (blue bar in the middle). The spin up (red) is transmitted, spin down (blue) is reflected. The electron-nuclear spin flip flops can happen at positions where an occupied initial state and empty final state with opposite spins exist. Therefore, the nuclear spin flop $\uparrow \to \downarrow$ can happen on the source side of the scatterer, the opposite one on the drain side. It leads to nuclear polarization (fat arrows) of equal magnitude and opposite sign on the opposite sides. Around the scatterer, the nuclear diffusion smoothens the polarization profile into a linear crossover over the distance $l_{\rm cross}$.
The polarization in the leads---of both electrons and nuclei---will decay to zero beyond the spin diffusion length away from the scatterer (this effect is reflected neither in this figure, nor in our model).
}
\end{figure}

We now discuss two cases of interest. First, assume that  the equilibration is dominated by other sources than the electrons flowing through QPC. We can then take the total rate $\Gamma \approx W^{(0)}$ as spatially uniform, and, for a scatterer with symmetric leads, we get,
\be
\dIzx{x} = - \dIzx{-x},
\label{eq:main1}
\ee
an anti-symmetric non-equilibrium dynamical nuclear spin polarization around the scatterer. The nuclear spin diffusion smearing the polarization allows us to extend the validity of this equation from $x$ in the leads to all $x$. The second case, if the equilibration through the QPC electrons is a substantial part of the total rate $\Gamma$, the latter is asymmetric on the two sides, and makes the nuclear spin polarization larger in the source compared to the drain. Again, due to the nuclear spin diffusion, around the QPC center one expects to find a surplus of nuclear spin polarization from the source lead. With the $g$ factor negative in GaAs, this will be the nuclear spin down,
\be
\dIzx{0} <0.
\label{eq:main2}
\ee
The symmetry relations in Eqs.~\eqref{eq:w2sym} and \eqref{eq:w1sym}, and their consequence for the polarization, Eqs.~\eqref{eq:main1} and \eqref{eq:main2}, constitute our main results.

\begin{figure}
\includegraphics[width=0.49\columnwidth]{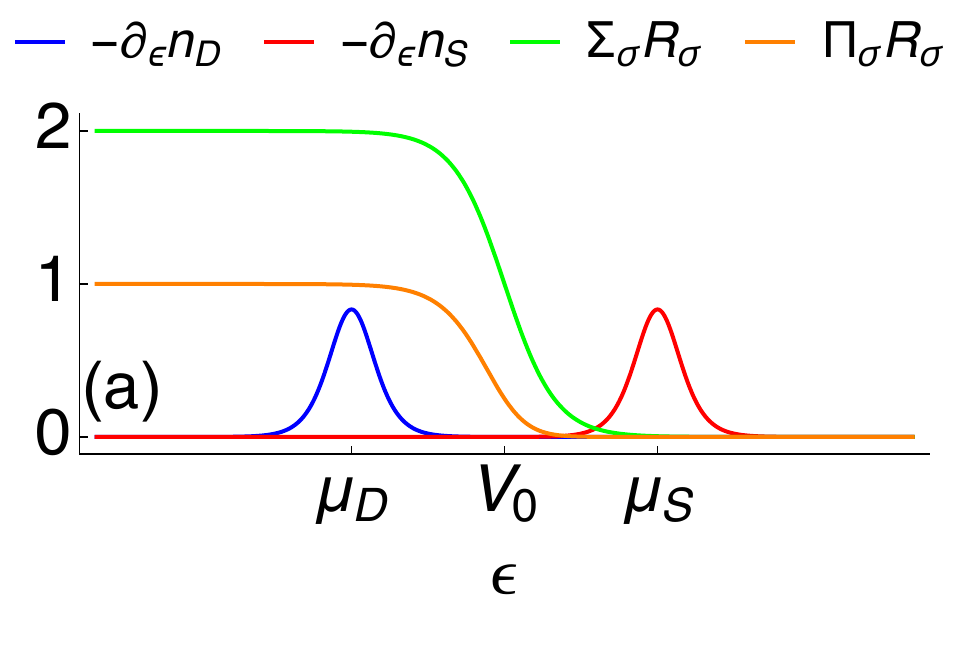}
\includegraphics[width=0.49\columnwidth]{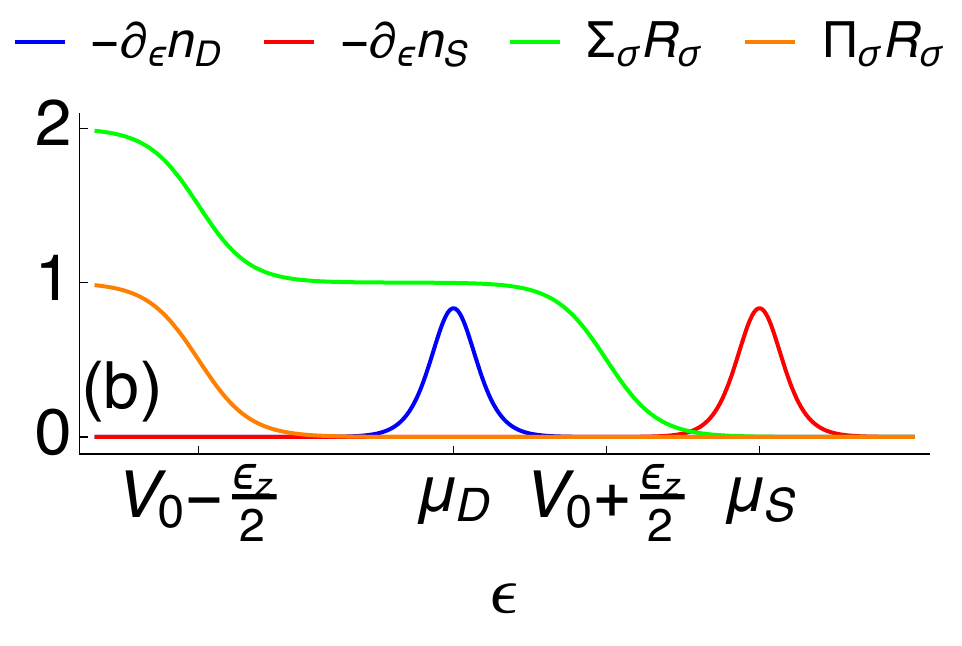}
\includegraphics[width=0.49\columnwidth]{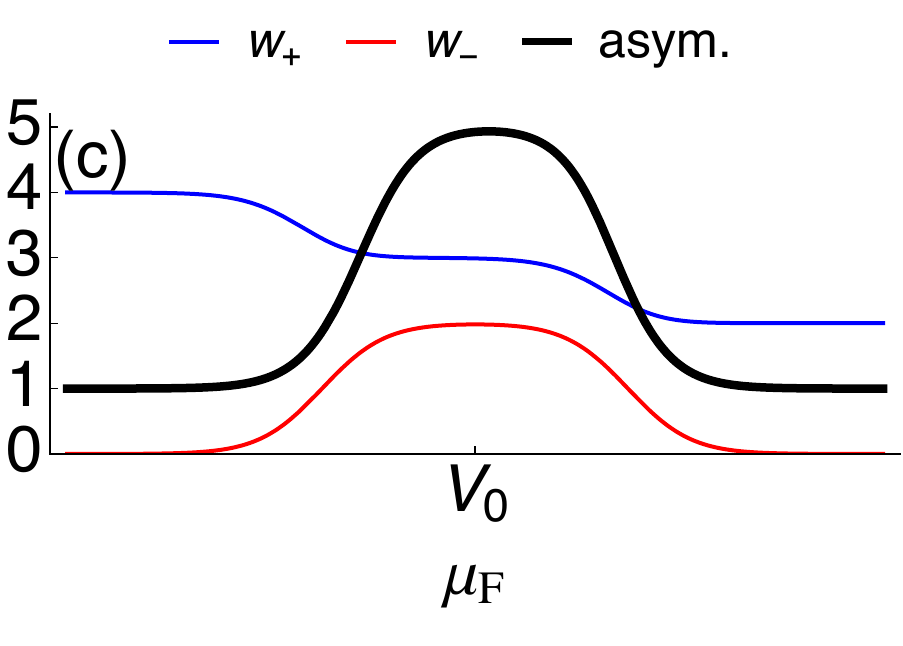}
\includegraphics[width=0.49\columnwidth]{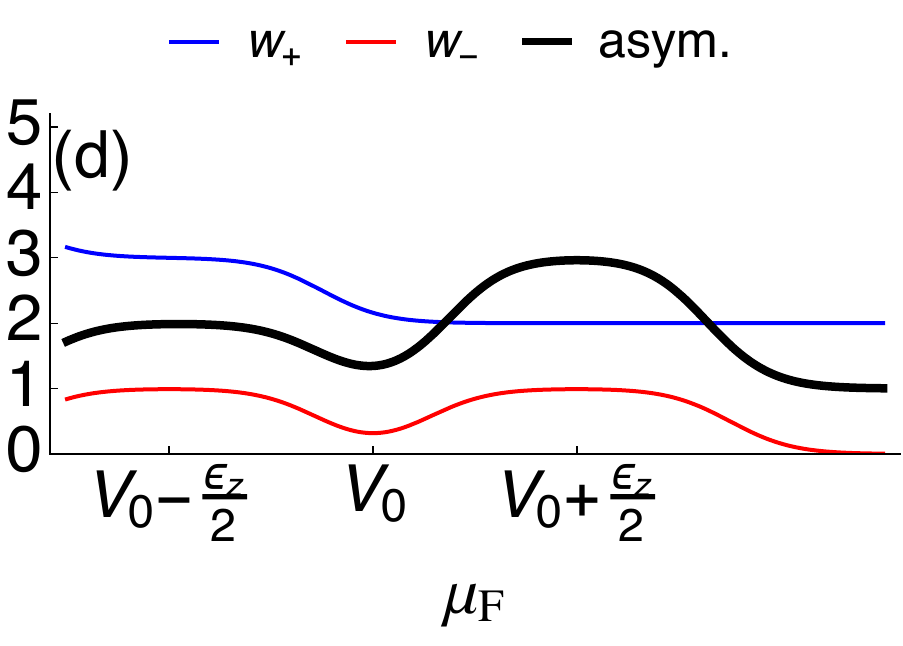}
\caption{\label{fig:equilibration}
Illustration of the equilibration asymmetry.  (a) Shows the energy dependence of the lead occupations derivative $-\partial_\epsilon n_l(\epsilon)$ and the sum and product of reflection probabilities $R_\sigma = R(\epsilon_{\rm kin}^\sigma)$, for $\epsilon_z \ll eV$ and $V_0=\mu_F$. (c) Shows the symmetric and anti-symmetric parts of the equilibration in the leads, $w_\pm = w(x_D) \pm w(x_S)$, according to Eq.~\eqref{eq:w1sym}, in common arbitrary units. The asymmetry factor $w(x_D)/w(x_S)$ is plotted in black. The maximal asymmetry is a factor of 5 and arises in the configuration shown in (a). (b) and (d) are analogous to (a) and (c) for $\epsilon_z = 8$, $eV=6$, $V_0=-2$, and [in (b)] $\mu_F=2$, in common arbitrary units. The maximal asymmetry is 3, corresponding to the configuration on (b).\\
}
\end{figure}

The physical origin of these relations is easy to understand. Concerning the symmetry of the polarization, consider an electron impinging on the scatterer from the source at an energy inside the bias window and such that the scattering is, for simplicity, fully spin selective, as illustrated in Fig.~\ref{fig:dipole}. The electron with spin $\uparrow$ is transmitted, with $\downarrow$ is reflected. Both incoming and outgoing waves in the source lead are thus occupied for spin $\downarrow$, while for the spin $\uparrow$, the incoming wave is occupied in the source lead and the outgoing wave is occupied in the drain lead. Under these conditions, in the source lead, the only spin-flip elastic scattering that the electron can make is $\downarrow \to \uparrow$, with two options concerning the propagation direction: incoming $\to$ outgoing, and outgoing $\to$ outgoing. In the drain lead, one finds also two options, this time 
outgoing $\to$ outgoing, and outgoing $\to$ incoming for the spin-flip scattering in the opposite direction. If the leads are identical, the surplus of the spin density created by the spin-selective scatterer in the occupied states in one lead is exactly compensated by the surplus of the unoccupied states in the other lead. This compensation is the reason for the exact symmetry relation in Eq.~\eqref{eq:main1}.
As we already stated, this result is not conditioned on the time-reversal symmetry, and is therefore valid also at strong magnetic fields. It will be also immune to the effects of electron-electron interactions, if they can be described as an effective mean-field potential,\footnote{The requirement is that the electron scattering can be still considered elastic. On the other hand, the effective potential can be an arbitrary function of the applied voltage and position, without any symmetry relations required.} 
included in the potential of the scatterer $V_Q(x)$.  
We therefore conclude that a spin-selective scatterer is generally accompanied by an electron spin density dipole, which results in the dipole-like spatial profile in the dynamically created polarization of nuclei. This is analogous to the charge density dipole accompanying a charge scatterer \cite{landauer1975:ZPB}.

Turning now to the asymmetry of the equilibration,
\footnote{By the asymmetry we mean the ratio $r=w_{\overline{\sigma}\sigma}^{(1)}(x_S) / w_{\overline{\sigma}\sigma}^{(1)}(x_D)$. It is given by $r=(A+B)/(A-B)$, where $A$ is the right-hand side of Eq.~\eqref{eq:w1sym2} and $B$ is the right-hand side of Eq.~\eqref{eq:w1sym1}.}
 we first note, e.g., looking at Eq.~\eqref{eq:wlimit1}, that the equilibration is related with the states emanating from a lead at its Fermi surface. Considering again a simplified situation, where the applied voltage is such that all states at the chemical potential of the source are transmitted, they contribute equally for the equilibration in the source and in the lead. The fully reflected states at the chemical potential of the drain do not reach the source, and since the electron density is doubled by the reflection, these states contribute by a factor 4 times bigger to the equilibration in the drain. The resulting asymmetry between the leads would be a factor of 5 (see the left column of Fig.~\ref{fig:equilibration}), a maximal possible asymmetry. Similar considerations in the opposite regime $eV \gg \epsilon_z$ show that in this case the maximal asymmetry is a factor of 3 (see the right column of Fig.~\ref{fig:equilibration}).

\section{Effects on QPC conductance: resistively detected NMR}

We now turn to possible experimental signatures of the dynamically created nuclear spin polarization analyzed in the previous section. To this end, we consider the resistively detected NMR, where the conductance through the QPC is monitored upon scanning the microwave frequency around one of the NMR frequencies of the material. On resonance, the NMR field couples to nuclear spins, adding $W^{(3)}$ among the rates entering Eq.~\eqref{eq:Iz1}. Since this rate fulfills $W_{\overline{\sigma}\sigma}^{(3)}=W_{\sigma\overline{\sigma}}^{(3)}$, in Eq.~\eqref{eq:Iz3} it contributes only to the depolarization $\Gamma$. The resulting decrease in nuclear polarization decreases the Overhauser field contribution to the total energy splitting of the spin opposite electrons. The essence is to tune the QPC such that it becomes sensitive to this change.

With the aim of confronting the theory against the data measured in Ref.~\onlinecite{kawamura2015:PRL}, we calculate the NMR response from our theory analytically, using a simple perturbation theory, and rely on numerics for more quantitative and robust statements. We begin with the analysis of the system without electron-electron interactions. As we will see below, this approach predicts a conductance change upon applying the NMR depolarization, $\Delta G$, with a sign opposite to what is observed in the experiment. This does not seem to be an issue of the adopted approximations, as numerics shows the same discrepancy. We then proceed to the interacting case where, using numerics, we indeed observe a sign reversal for strong enough electron-electron interactions. With this outlook for the section contents, we now proceed to details. 

\subsection{Expected NMR conductance signal}

In the experiment, the nuclear polarization and its detection proceed under separate conditions. The first is done by applying a large current through the QPC, so that $eV \gtrsim k_B T$. As described in Sec.~\ref{sec:dipole}, it results in a build-up of the non-equilibrium polarization $\dIzx{x}$ according to Eq.~\eqref{eq:Iz4}. It is spatially dependent, mainly due to the spatial dependence of the pumping rate, given in Eq.~\eqref{eq:W2}. The detection is done under a much lower bias $eV \ll k_B T, \epsilon_z$, at possibly a different value of the gate voltage, chosen to maximize the signal. In this regime, the current $\mathcal{I}$ through the QPC is contributed by the two spin species,
\be
\mathcal{I} = \frac{e}{2\pi \hbar} \int {\rm d}\epsilon \sum_\sigma T_\sigma(\epsilon) [n_S(\epsilon) - n_D(\epsilon) ],
\label{eq:current}
\ee
according to the spin-resolved transition probabilities $T_\sigma(\epsilon) =T(\epsilon_{\rm kin}^\sigma)$. Because of the small bias, we can assume linear response, where the differential conductance $G = \partial_V \mathcal{I}$ is
\be
G = -\frac{e^2}{2\pi \hbar} \int {\rm d}\epsilon \sum_\sigma  T_\sigma (\epsilon)  \partial_\epsilon n(\epsilon).
\label{eq:G}
\ee
The nuclear spin polarization influences this conductance through the induced Overhauser field as a position-dependent contribution to the Zeeman energy
\be
\delta \epsilon_z(x) = A \int {\rm d}y {\rm d}z \rho(x,y,z) \dIzx{x,y,z} = A' (x) \dIzx{x}.
\label{eq:dez}
\ee
We used Eqs.~\eqref{eq:W2} and \eqref{eq:Iz4}, to perform the integration over the transverse coordinates, and introduced an effective cross section
\be
A'(x) = A \times \frac{\int {\rm d}y {\rm d}z | \phi(x,y,z)|^6}{ | \phi(x,0,0)|^4 },
\label{eq:Ap}
\ee
and we also use $A^\prime(0) \equiv A^\prime$ in what follows.

The change of the conductance upon nuclei depolarization by microwave field constitutes the signal extracted experimentally. Assuming that the NMR pulse decreases the nuclear polarization to some fraction $1-p \in [0,1]$ from its initial value $\dIz$, the signal is
\be
\Delta G =G\{ (1-p) \dIz \} -  G\{ \dIz \},
\label{eq:DG}
\ee
where $p$ depends on the NMR field amplitude and duration.
In this equation, the conductance is a functional of the dynamical nuclear polarization. Since it is difficult to evaluate this equation analytically exactly even in the simplest cases, we do it numerically. We adopt a model which includes the electron-electron interactions, as described in App.~\ref{app:num}. To understand the results obtained from the model, we also perform some rough estimates.

\subsection{Signal for a constant polarization}

\begin{figure}
\includegraphics[width=\columnwidth]{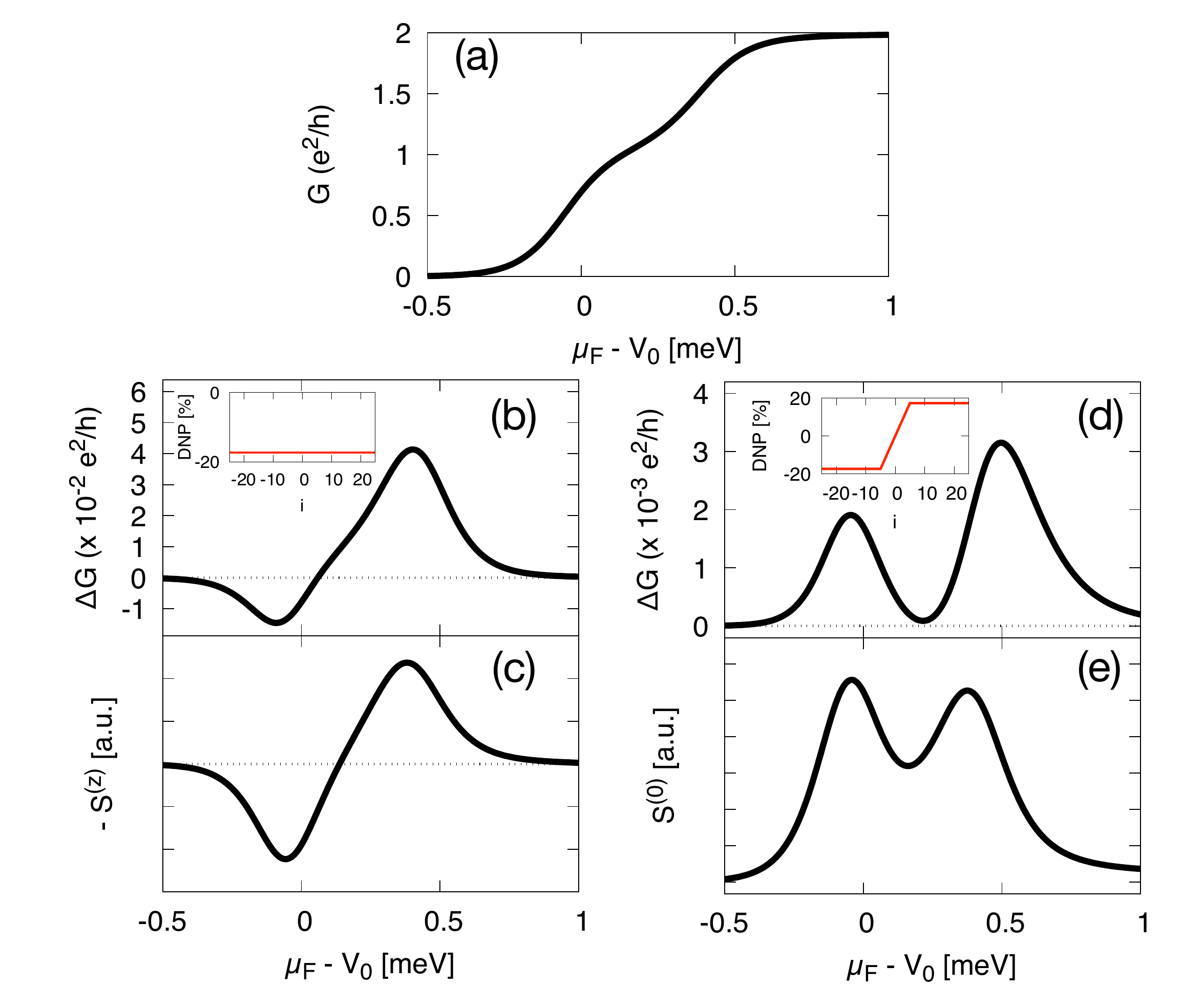}
\caption{\label{fig:small_U}
Illustration of the signal for different nuclear spin polarization patterns. In all panels, we plot quantities as a function of the gate voltage $V_0$ measured relative to the Fermi energy $\mu_{\rm F}$. We take parameters comparable to the experiment of Ref.~\onlinecite{kawamura2015:PRL}: $B=4.5$ T, $\hbar \omega =$ 0.5 meV, and $T=0$ for simplicity. The system is interacting, with the Coulomb energy parameter $U=6.4$ meV (see App.~\ref{app:num}). (a) The conductance $G$. Note that a finite value of the electron-electron interaction parameter is necessary to produce a noticeable conductance plateau at $G=e^2/h$, as observed in the experiment. 
(b) $\Delta G$ calculated according to Eq.~\eqref{eq:DG} for $p=0.5$ and a uniform initial nuclear polarization (shown in the inset). (c) The spin shot noise, Eq.~\eqref{eq:Sspin}. (d) $\Delta G$ calculated according to Eq.~\eqref{eq:DG} for $p=0.5$ and a dipole-like nuclear polarization (shown in the inset). (e) The charge shot noise, Eq.~\eqref{eq:Scharge}.
}
\end{figure}

To proceed, we assume that Eq.~\eqref{eq:DG} can be linearized (the Overhauser effects on the electron scattering, and/or its NMR induced change, are small),
\be
\Delta G = - p \int {\rm d}x\, \frac{\partial G}{\partial \dIzx{x}}{\dIzx{x}},
\ee 
and approximate the complicated functional dependence of the conductance by the sensitivity to the energy at the QPC center only, $\partial G/\partial \dIzx{x} \propto \delta(x)$. We get
\be
\Delta G \approx p  \dIzx{0}  A^\prime \partial_{\epsilon} \left( \frac{e}{2 \pi \hbar} \sum_\sigma \sigma \,  T_\sigma( \epsilon) \right)|_{\epsilon=\mu_{\rm F}}.
\label{eq:DG1}
\ee 
The signal is proportional to the nuclear polarization at the QPC center and the energy derivative of the spin-polarized transmission probability, in other words, to the derivative of the spin current in the QPC. Using the specific conductance of Eq.~\eqref{eq:TQPC}, which gives
\be
\partial_{\epsilon} T(\epsilon) = \frac{2\pi}{\hbar \omega} T(\epsilon) [1-T(\epsilon)],
\label{eq:deT}
\ee
we can also write
\be
\Delta G \approx p  \dIzx{0} \frac{A^\prime}{\hbar \omega} \frac{e}{\hbar} S^{(z)},
\label{eq:DG2}
\ee 
which states that the signal is proportional to the spin polarization of the current noise,
\be
S^{(z)} = \sum_\sigma \sigma \, T_\sigma(\mu_{\rm F}) \left[ 1- T_\sigma(\mu_{\rm F}) \right].
\label{eq:Sspin}
\ee
These formulas predict a signal which has opposite sign on the flanks of the two spin-resolved plateaus. Assuming a negative nuclear polarization at the QPC center, $\dIzx{0}<0$, following Eq.~\eqref{eq:main2}, $\Delta G$ is negative for $G < e^2/h$ and positive for $G>e^2/h$. 
Figures \ref{fig:small_U}(b) and \ref{fig:small_U}(c) illustrate this case.

In the experiment of Ref.~\onlinecite{kawamura2015:PRL}, the NMR response was measured on the flank of the spin down plateau, $G>e^2/h$, where a robust negative signal was found, $\Delta G <0$. This is the opposite behavior that we obtain from Eq.~\eqref{eq:main2}. However, as already noticed in the above, the nuclear spin equilibration in that experiment is most probably dominated by other channels than electrons, so that Eq.~\eqref{eq:main1} applies for the profile of the dynamical nuclear polarization. The latter, however, leads to $\dIzx{0}=0$, and a zero response follows from Eq.~\eqref{eq:DG2}.

\subsection{Signal for a dipole-like polarization pattern}
To estimate the NMR response in the case of the dipole-like dynamical nuclear polarization, one has to go beyond the approximations adopted to arrive at Eq.~\eqref{eq:DG1}. We will do it in the following way. 
We first linearize the Overhauser field around the QPC center
\be
\dIzx{x} \approx x \frac{\dIzx{x_D}-\dIzx{x_S}}{2l_\textrm{cross}},
\label{eq:dipolelin}
\ee
where $\dIzx{x_l}$ is the nuclear polarization in lead $l$, and $l_\textrm{cross}$ is the length over which the polarization crosses from the saturated value at the source lead, $\dIzx{x_S}$, to its saturated value in the drain lead, $\dIzx{x_D}=-\dIzx{x_S}$. The lead polarizations can be obtained from Eq.~\eqref{eq:Iz4}, evaluating Eq.~\eqref{eq:W1} at the corresponding lead putting $x_n=x_{S/D}$. Assuming the QPC is gated to be spin selective at the Fermi energy, we get an order-of-magnitude estimate ($V>0$)
\be
\dIzx{x_D} \sim \frac{4}{3}I(I+1) W_0 \frac{\textrm{min}(eV,\epsilon_z)}{\hbar \Gamma}.
\label{eq:only}
\ee
The crossover length, on the other hand, is approximately the sum of the QPC effective width $l_{\rm QPC}$ and the nuclear spin diffusion length $\surd{D/\Gamma}$, with $D$ the nuclear spin diffusion coefficient.\footnote{Taking a typical value $D=7$ nm s$^{-1}$ \cite{gong2011:NJP, deng2005:PRB}, and $\Gamma^{-1} = 100$ s estimated in Ref.~\onlinecite{kawamura2015:PRL}, we get the diffusion length of 26 nm.}

The linearized Overhauser energy can be included simply into the scattering potential. Indeed, using Eqs.~\eqref{eq:VQPC} and \eqref{eq:dipolelin} we get
\be
V_Q(x) + \sigma \delta \epsilon_z(x) = V_Q( x - \sigma \xi l_{\rm QPC}) + \frac{1}{2}\hbar \omega \xi^2,
\label{eq:VSdipole}
\ee
that is, a spin-dependent coordinate shift, and an overall constant. Both are proportional to the dimensionless factor
\be
\xi = \frac{l_{\rm QPC}}{l_\textrm{cross}} \frac{A^\prime \dIzx{x_D} }{\hbar \omega},
\label{eq:xi}
\ee
which is small since $l_\textrm{cross}\leq l_{\rm QPC}$, and typically $A^\prime \ll \hbar \omega$. Since the coordinate shift does not change the transmission probability, we conclude that the influence of the dipole-like nuclear polarization on the conductance is obtained by the replacement $V_0 \to V_0 +\xi^2 \hbar \omega/2$ in Eq.~\eqref{eq:TQPC}. From here we immediately get
\be
\Delta G \approx \frac{1}{2}\xi^2 p(2-p) \hbar \omega \partial_{\epsilon} \left( \frac{e}{2 \pi \hbar} \sum_\sigma \,  T_\sigma( \epsilon) \right)|_{\epsilon=\mu_{\rm F}},
\label{eq:DG3}
\ee 
or, using Eq.~\eqref{eq:deT}, an equivalent expression follows, 
\be
\Delta G \approx \frac{1}{2} \xi^2 p(2-p) \frac{e}{\hbar} \,  S^{(0)}.
\label{eq:DG4}
\ee 
The most apparent difference is that now the NMR signal is proportional to the sum (rather than difference) of the current noises of the two spin species,
\be
S^{(0)} =  \sum_\sigma T_\sigma( \mu_{\rm F}) \left[ 1- T_\sigma( \mu_{\rm F}) \right].
\label{eq:Scharge}
\ee
The conductance change will therefore have the same sign on both flanks of the spin resolved plateaus, $\Delta G>0$. Figures \ref{fig:small_U}(d) and \ref{fig:small_U}(e) illustrate this case. The qualitative difference of the conductance change as a function of the gate potential can therefore distinguish the uniform and dipole-like nuclear polarization.

As a final remark, we point out the scaling of the signal with the basic parameters of the problem: the electron Zeeman energy does not directly enter to Eq.~\eqref{eq:xi} or Eq.~\eqref{eq:DG4}. It means that the RD-NMR signal is not directly dependent on the electron $g$ factor. There is an indirect influence: a larger $g$ factor allows a higher spin-polarized electronic current, as seen in Eq.~\eqref{eq:only}, the square of which the RD-NMR signal is proportional to. On the other hand, there is strong dependence on the nuclear spin length and especially the electron-nuclear coupling strength, $\Delta G \sim I^4 A^6 /\Gamma^2$. Assuming that the nuclear relaxation rate $\Gamma$ is not due to electrons, it will be independent on $A$. We therefore expect the electron-nuclear coupling $A$ to be the most important parameter for the overall magnitude of the effect.

\section{Signal sign reversal}

The results of the previous section predict a positive RD-NMR signal, $\Delta G >0$, at and beyond the 1/2 conductance plateau, $G>e^2/h$. This is so for both uniform, and a dipole-like nuclear polarization around the QPC center. This picture is confirmed by the numerical modeling (see App.~\ref{app:num}), with which we evaluate directly Eq.~\eqref{eq:DG}, without any further assumptions, and include also the effects of the electron-electron interactions. From extensive numerical investigations (not shown), we conclude that this prediction is robust with respect to the model details. Nevertheless, the experiment in Ref.~\onlinecite{kawamura2015:PRL} showed the opposite, $\Delta G<0$, slightly beyond the 1/2 conductance plateau. We now discuss the possible origins of the discrepancy.

\subsection{Due to strong electron-electron interactions}

\begin{figure}
\includegraphics[width=\columnwidth]{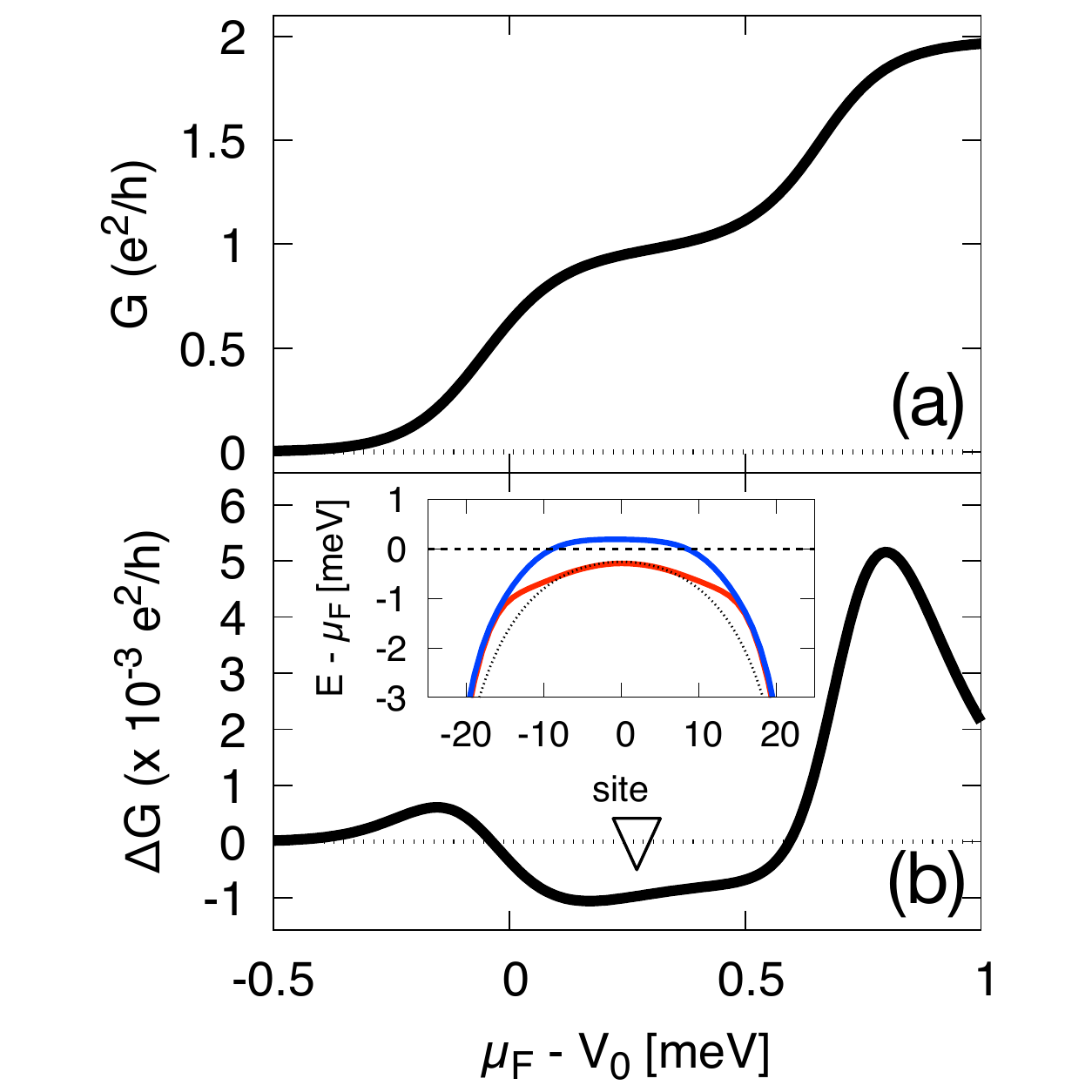}
\caption{\label{fig:large_U}
(a) $G$ and (b) $\Delta G$ as a function of $V_0$ for $U = 7.7$ meV. The remaining parameters are the same as those given in Fig.~\ref{fig:small_U}, with the nuclear polarization as in (e) therein. Inset in (b): the effective scattering potential $V_\sigma(x)$, including the mean-field interaction correction, as seen by the up (red curve) and down (blue curve) electron spins, at the gate voltage indicated by the triangle. The dotted line shows the bare potential [Eq.~\eqref{eq:VQPC}] and the dashed line is the Fermi energy. }
\end{figure}

During the mentioned numerical investigations, we found that a signal sign reversal appears for a strong Coulomb interaction. Figure \ref{fig:large_U} shows a typical example. The strong (repulsive) electron-electron interactions first of all pronounce the half-conductance plateau (widening the gate voltage interval where it is observed) compared to Fig.~\ref{fig:small_U}(a). Figure \ref{fig:large_U}(b) shows that in the regime where the electron-electron exchange spin polarizes the electrons in the QPC, the RD-NMR signal becomes negative, $\Delta G<0$.  Again, this is a robust feature, not due to parameters' fine tuning (see Fig.~\ref{fig:dg-neg-other} in App.~\ref{app:DGneg}). Even though the particular mechanism is not obvious, the negative sign is associated with clear deformations of the scattering potential---the inset of Fig.~\ref{fig:large_U} (b) illustrates how it departs from the bare inverted parabolic potential near the QPC center. We find it rather surprising that such seemingly non-generic changes do give rise to a robust negative signal. We also note that strong versus weak Coulomb interactions do not induce any qualitative changes for the signal from a uniform nuclear polarization, see Fig.~\ref{fig:large_U_uni} of App.~\ref{app:DGneg}.

The Coulomb interaction effects are essential to explain the 1/2 conductance plateau in the experiment in Ref.~\onlinecite{kawamura2015:PRL}. Indeed, the width of the observed plateau was much closer to Fig.~\ref{fig:large_U} (a) than to Fig.~\ref{fig:small_U} (a). It suggests that the electron-electron interactions could cause the signal sign reversal. On the other hand, we were not able to match that experimental data quantitatively: the numerical value for $\Delta G$ that we get is an order of magnitude smaller than the observed one. 
We therefore remain inconclusive, as of whether it was the interaction effects responsible for the negative signal observed in the experiment of Ref.~\onlinecite{kawamura2015:PRL} and leave it an open question for further experimental and numerical investigations.  

\subsection{Due to non-equilibrium spin polarization in the leads}

\label{sec:non-eq}

\newcommand{\dmu}[1]{\delta \mu_{#1}}

We now consider a situation with non-equilibrium spin polarizations in the leads \cite{rasly2017:PRB}. The latter corresponds to having spin-dependent chemical potentials 
\be
\mu_{\sigma l} \equiv \mu_l+\sigma \dmu{l},
\ee
with the ``spin voltage'' $\dmu{l}$ parametrizing the non-equilibrium electronic spin accumulation in the lead $l$ \cite{meair2011:PRB}. 
In this case, we rewrite the occupations from Eq.~\eqref{eq:n} as
\be
n_{l\sigma}(\epsilon) = n_l(\epsilon) + \sigma \dmu{l} \, \delta n_l(\epsilon).
\label{eq:dn}
\ee
For a spin voltage smaller than the temperature, the lowest-order Taylor expansion gives
\be
\delta n_l(\epsilon) = -\partial_\epsilon n_l(\epsilon) \xrightarrow{T \to 0} \delta(\epsilon-\mu_l),
\ee
where the last property holds in the limit $T \to 0$. We note that, however, the functions $\delta n_l(\epsilon)$ defined by Eq.~\eqref{eq:dn} are non-negative for any parameters' values, since that property relies only on the monotonicity of the Fermi distribution. 

With this expansion of the model, we now reinstate the spin voltages into Eq.~\eqref{eq:Wss}. Neglecting the nuclear Zeeman energy, for the rate rescaled according to Eq.~\eqref{eq:wss} we get 
\be
w_{\overline{\sigma}\sigma} =  \sum_{l l^\prime} \int \frac{{\rm d}\epsilon}{\hbar} |\Psi_{\epsilon l \sigma}|^2|\Psi_{\epsilon l^\prime \overline{\sigma}}|^2  n_{l\sigma}(\epsilon)  [1 - n_{l^\prime \overline{\sigma}}(\epsilon) ],
\ee
where we again omitted the position argument $x_n$ of the rate and the wave functions. Using here Eq.~\eqref{eq:dn}, we get the change in the nuclear spin-flip rate due to the finite spin voltages as
\be
\begin{split}
\delta w_{\overline{\sigma}\sigma} & =  \sum_{l l^\prime} \int \frac{{\rm d}\epsilon}{\hbar} |\Psi_{\epsilon l \sigma}|^2|\Psi_{\epsilon l^\prime \overline{\sigma}}|^2
 \Big\{ \sigma \, \dmu{l} \, \delta n_{l}(\epsilon)  [1 - n_{l^\prime}(\epsilon) ] \\
&-\overline{\sigma} \, \dmu{l^\prime} \, \delta n_{l^\prime}(\epsilon)  n_{l}(\epsilon) - \sigma \overline{\sigma} \, \dmu{l} \, \dmu{l^\prime}  \, \delta n_{l}(\epsilon) \delta n_{l^\prime}(\epsilon) \Big\}.
\end{split}
\ee
The last term in curly brackets is symmetric upon inverting the spin indexes, and therefore contributes only to the equilibration. We denote it by $\delta w^{(I)}_{\sigma \overline{\sigma}}$, using analogous superscript notation as in the previous section. The remaining terms represent the polarization part, 
\be
\begin{split}
\delta w^{(II)}_{\overline{\sigma}\sigma} & =  \sigma \sum_{l l^\prime} \int \frac{{\rm d}\epsilon}{\hbar} |\Psi_{\epsilon l \sigma}|^2|\Psi_{\epsilon l^\prime \overline{\sigma}}|^2 \\
& \times \Big\{ \dmu{l}\, \delta n_{l}(\epsilon)  [1 - n_{l^\prime}(\epsilon) ]
+ \dmu{l^\prime} \, \delta n_{l^\prime}(\epsilon)  n_{l}(\epsilon) \Big\}.
\end{split}
\label{eq:non-eq-main}
\ee
This equation constitutes the main result of this section. It shows that finite spin voltages induce an additional polarization, the sign of which is given by the sign of the spin voltage: a positive spin voltage in a lead, $\dmu{l}>0$, meaning a surplus of spin-up electrons, induces a flow of spin up into the nuclear ensemble. One can see this from Eq.~\eqref{eq:non-eq-main}, 
noting that the functions $n$, $1-n$, $\delta n$, and $|\Psi|^2$ are non-negative. 

For illustration, we also consider the limit $|\dmu{l}| \ll k_B T \ll |e V|, \hbar \omega$, which gives
\begin{subequations}
\begin{eqnarray}
\delta w^{(I)}_{\overline{\sigma}\sigma}(x_n) &=&  \sum_l \frac{(\dmu{l})^2}{6\hbar  k_B T} |\Psi_{\mu_l l \sigma}(x_n)|^2 |\Psi_{\mu_l l \overline{\sigma}}(x_n)|^2,\phantom{xxx}\\
\delta w^{(II)}_{\overline{\sigma}\sigma}(x_n) &=&  \sigma \sum_{l l^\prime} \frac{\delta \mu_l}{\hbar} |\Psi_{\mu_l l \sigma}(x_n)|^2 |\Psi_{\mu_l l^\prime \overline{\sigma}}(x_n)|^2,
\end{eqnarray}
\end{subequations}
where we reinserted the position argument explicitly. These equations demonstrate the general properties of the equilibration and polarization rate contributed by finite spin voltages: the dominant effect of non-equilibrium electronic spin polarization is a transfer of this polarization into nuclei. In addition, there is a much smaller increase in the equilibration rate.

\section{Dispersive lineshape}

\begin{figure}
\includegraphics[width=\columnwidth]{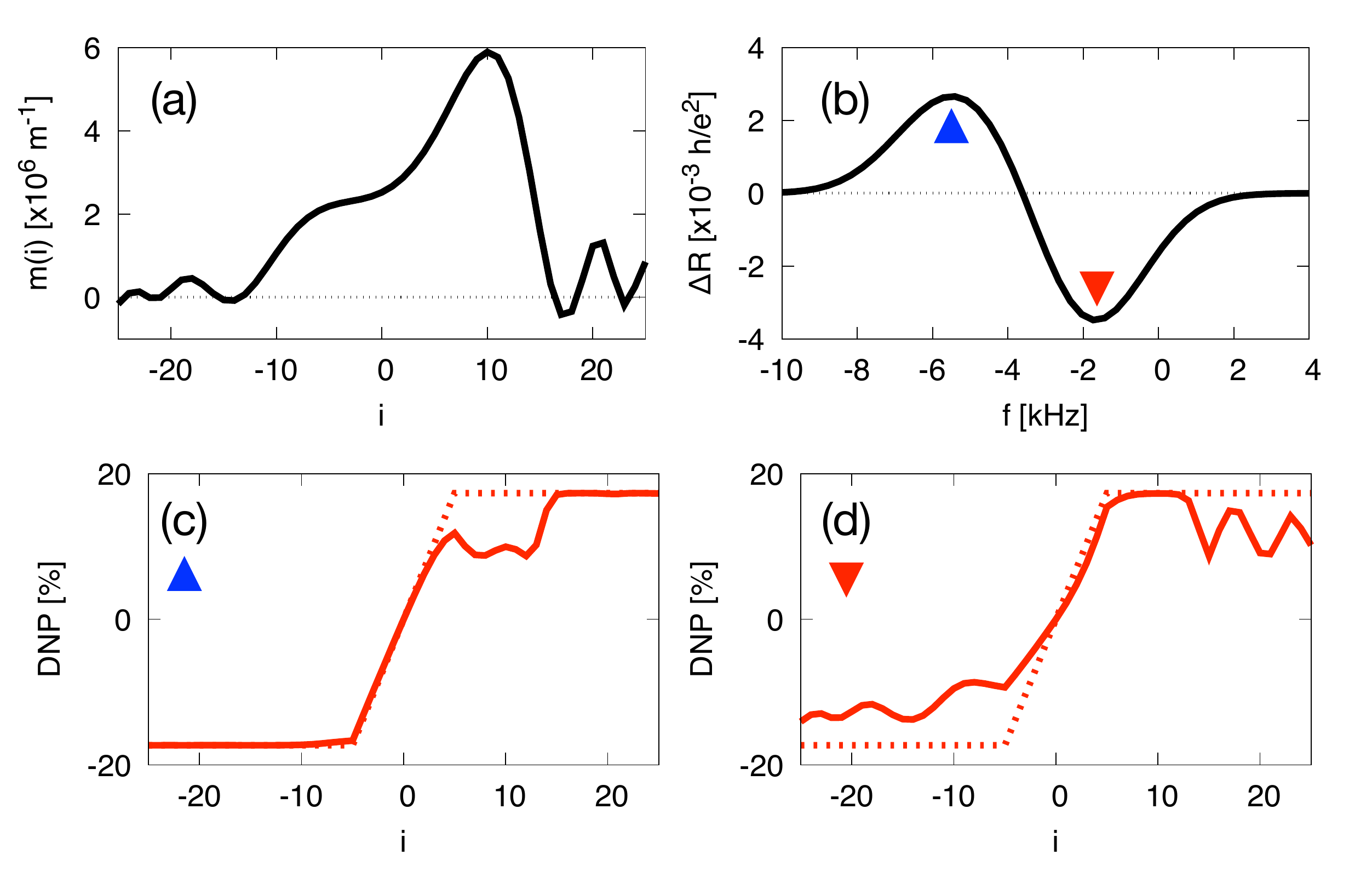}
\caption{\label{fig:rdnmr}
The RD-NMR signal at a finite bias, $e V = 0.3$ meV, and $\mu_{\rm F}- V_0 = 0.2$ meV. The output of the numerical model described in App.~\ref{app:num} is plotted. (a) The electron spin-polarization density $m(i) = \left\langle  n_{\uparrow}(i) - n_{\downarrow}(i)  \right\rangle/ a $ as a function of the site index $i$. (b) $\Delta R$ as a function of the NMR frequency $f$ calculated according to $R=V/I$, and Eqs.~\eqref{eq:dR} and \eqref{eq:pxf}. (c), (d)  The DNP rate before (dotted line) and after (solid line)
the depolarization by NMR, for frequency denoted by the respective colored triangle in (b). Apart from the bias, the parameters are the same as in Fig.~\ref{fig:small_U} except for the temperature $T=10$ mK, adopted to regularize the step functions in the occupations for reasons related to numerics.
}
\end{figure}

In the previous section, we considered the RD-NMR signal detected at zero bias voltage. In that case, the electronic spin density, even though position dependent, is spatially symmetric around the QPC. The associated Knight field is then also symmetric, and it is reasonable to assume that the NMR field depolarizes the nuclei uniformly. Here, we consider effects at finite bias. Surprisingly, we find that our theory of the nuclear dipole-like polarization predicts a RD-NMR signal with a ``dispersive line shape'', which has been observed in several experiments in two-dimensional electron gas (2DEG) in the quantum Hall regime \cite{desrat2002:PRL,stern2004:PRB,gervais2005:PRB,kodera2006:PRB,tracy2006:PRB,dean2009:PRB,bowers2010:PRB,yang2011:APL,desrat2013:PRB,desrat2015:JPCM}. It refers to the RD-NMR signal which, as a function of the NMR frequency, resembles the shape of a derivative of a Lorentzian curve, similar to the one plotted on Fig.~\ref{fig:rdnmr}(b).  

The origin of such signal here is the following. At a finite bias, the non-equilibrium electronic spin accumulation created by the QPC gives rise to the dipole-like nuclear polarization, as explained in Sec.~II. If, contrary to the assumptions in Sec.~III, the detection by the NMR field is done also at a finite bias, the Knight field for the nuclei is asymmetric across the QPC. The nuclei in the drain see more electrons with spin up, and therefore their NMR resonance frequency is smaller, compared to nuclei in the source. If the difference is large enough, the two sets of nuclei are depolarized selectively. Since these two sets of nuclei are polarized oppositely, their depolarization is expected to lead to roughly opposite change in the conductance.

To confirm this qualitative analysis, we resort to numerics. Adopting the same model as before (see App.~\ref{app:num}), we get that a finite bias indeed leads to an asymmetric electron spin accumulation around the QPC, as shown in Fig.~\ref{fig:rdnmr}(a). We note that electron-electron interactions are essential to achieve a substantial asymmetry. 
To comply with the experiments reporting on the dispersive line shape, the RD-NMR signal in the resistance $R=V/\mathcal{I}$  is considered,
\be
\Delta R = R\{ ( 1 - p(x, f) ) \langle    \delta I(x) \rangle\} - R\{ \left\langle  \delta I(x)  \right\rangle\},
\label{eq:dR}
\ee
where, to include the position-dependent Knight shift, we promote the depolarization factor $p$ to a position- and frequency-dependent function,
\be
p(x,f) = p_0 \exp \left(- \frac{[f -f_0-\alpha m(x) S_\perp^{-1}]^2}{2 \gamma^2}  \right).
\label{eq:pxf}
\ee
Here, $p_0$ is the overall depolarization scale, $f$ is the frequency of the NMR field, $f_0$ is the resonance frequency of the given isotope, and $m(x) = n_\uparrow(x)-n_\downarrow(x)$ is the one-dimensional electron spin density. Equation \eqref{eq:pxf} was used to fit experimental data in Ref.~\onlinecite{kawamura2015:PRL}, showing good agreement for realistic parameters $\alpha =  -2.1 \times 10^{-22}$ kHz m$^3$ (for As atom), and $\gamma = 1.36$ kHz. Using it here, the calculated resistance shown in Fig.~\ref{fig:rdnmr}(b) displays a clear dispersive line shape as a function of frequency. Panels (c) and (d) confirm that at frequencies corresponding to the two different line shape extrema, the nuclei on opposite sides of the QPC are depolarized predominantly. This constitutes the main result of this section.

Before moving on, a comment is in place. In the experiments with 2DEG in the quantum Hall regime, the dispersive line shape (we denote it QHE-DL) was originally conjectured to signal the presence of a skyrmion crystal \cite{desrat2002:PRL}. This was later refuted, as the same feature was observed in parameters regimes where a skyrmion crystal is highly improbable \cite{gervais2005:PRB, bowers2010:PRB}. The current understanding is that the peak and dip are unrelated \cite{stern2004:PRB, kodera2006:PRB}, arising due to the response of nuclei positioned at different electronic states (spin unpolarized and polarized) \cite{desrat2013:PRB}. Even though we believe that a dispersive line shape connected to the nuclear dipole around a spin-sensitive scatterer (we denote it dipole-DL) is very general, we do not suggest that it directly offers an explanation of the mentioned 2DEG experiments, because of the following differences. Most importantly, the dipole-DL requires non-uniform, and therefore non-equilibrium nuclear spin polarization. The QHE-DL is on the other hand observed also for uniformly polarized nuclei (thermal polarization), at a rather small current \cite{desrat2002:PRL}, or without a dependence on the current \cite{tracy2006:PRB}. (On the other hand, the importance of the DNSP for for QHE-DL was pointed out in Refs.~\onlinecite{stern2004:PRB,dean2009:PRB}). Second, the dipole-DL in GaAs is expected to show the dip at a frequency smaller than the peak (see Fig.~\ref{fig:rdnmr}b). Such QHE-DL have been seen too \cite{yang2011:APL,desrat2015:JPCM,fauzi2017:PRB}, but the opposite shape is perhaps more standard \cite{desrat2002:PRL}.

\section{Conclusions}

We investigated the dynamical nuclear polarization arising at a quantum point contact gated to conductance $e^2/h$ in strong magnetic field. Our main message is that such a spin-selective scatterer gives rise to a local imbalance of the electronic spin polarization, which is transferred into nuclear spins, as a spatially asymmetric polarization pattern. 

To understand the pattern origin, it is useful to consider a simple example of a spin-filtering QPC, which, within the bias window, reflects the spin down electrons and transmits the spin up ones. This spin-dependent scattering creates a local non-equilibrium electronic spin, for parameters of GaAs, down on the source side and up on the drain side of the QPC. The actual polarization of the nuclear spins happens in these regions, somewhere between the QPC potential top and the leads (defined as where the electrons are in equilibrium, including their spin). The same pattern is then imprinted into the nuclear spin polarization, spin down on the source side, and spin up on the drain side. We have denoted these regions of substantial nuclear polarization on Fig.~\ref{fig:scatterer} by the pairs of the vertical dashed lines. Interested in the nuclear polarization close to a symmetric QPC, the nuclear polarization is exactly anti-symmetric around its potential top, with zero DNSP at the QPC center.

To produce a net nuclear polarization around the QPC center, an additional asymmetry is therefore necessary. An obvious possibility is some geometrical asymmetry of the scatterer, e.g., a difference of the source and drain sides. The characteristic feature of it is that the net polarization should swap upon inverting the bias voltage polarity, providing a very simple criterion straightforwardly testable in experiments. We identify an interesting additional possibility, an asymmetry connected to the nuclear spin dissipation. We find that the electronic contribution to it (the "Korringa relaxation") is typically substantially larger on the drain side, and therefore can lead to a net overall nuclear spin polarization with the sign given by the non-equilibrium spin on the source side (electron/nuclear spin down in GaAs). However, since usually the spin-lattice and nuclear diffusion dominate the electronic contribution to the nuclear spin relaxation at low temperatures, we predict that the nuclear dipole is the most typical situation to be expected. 

The two scenarios (dipole-like versus uniform DNSP) result in qualitatively different resistively detected NMR signals, both as a function of the QPC conductance, and as a function of the NMR frequency. For the former, the RD-NMR signal is similar to the spin and charge current noise (see Fig.~\ref{fig:small_U}), for the uniform and dipole-like nuclear polarization, respectively. For the latter, the dipole-like nuclear polarization leads to a dispersive line shape. We have confronted our theory to the experiment of Ref.~\onlinecite{kawamura2015:PRL}, but found that more data would be needed to confirm our predictions. This issue has to do with the detection of the nuclear dipole, which is complicated by the supposedly dominant role of the electron-electron interactions on the minute changes of the QPC conductance. 

\acknowledgments

We would like to thank K. Ono and T. Komine for useful discussions. This work was partly supported by the MEXT Grant-in-Aid for Scientific Research on Innovative Areas "Science of hybrid quantum systems", Grant No.~16H01047 and 15H05867. P. S. would like to thank for the support from JSPS Kakenhi Grant No.~16K05411 and CREST JST (JPMJCR1675).

\appendix

\section{Numerical model} 
\label{app:num}

To include the effects of electron-electron interactions, we model the QPC by a 1D tight-binding Hamiltonian,
\begin{equation}
	H = \sum_{j, \sigma} \epsilon_{\sigma}(j) c_{j, \sigma}^{\dagger} c_{j, \sigma} 
		- t \sum_{j, \sigma} c_{j, \sigma}^{\dagger} c_{j+1, \sigma} 
		+ \sum_{j}U_j n_{j, \uparrow} n_{j, \downarrow}.
\end{equation}
Here, $c_{j, \sigma}^{\dagger}$ creates an electron with spin $\sigma$
at the $j$-th site ($-N \leq j \leq N$) of the tight-binding chain which has a hopping amplitude $t=12.8$ meV, the nearest-neighbor distance $a=6.67$ nm, and $N=25$. The QPC potential energy and the Zeeman energy are included in the on-site energy,
$\epsilon_{\sigma}(j) = \epsilon(j) +\sigma  E_{\rm Z}(j)$.
We adopt the following potential:
\begin{equation}
\epsilon(j) = V_0 \exp \left( -\frac{(\hbar \omega j)^2}{4 V_0} \frac{1}{1-(j/N)^2}  \right),
\end{equation}
which smoothly connects the inverted parabola near the QPC  center ($j = 0$) with a constant in the leads $V_S=0=V_D$.
The Zeeman energy is $E_{\rm Z}(j) = g \mu_{\rm B} B(j)/2$,
with the total magnetic field
\begin{equation} \label{eq:}
 B(j) = B_{\rm ex} + B_{\rm N}(j),
\end{equation}
contributed by the external field $B_{\rm ex}$ and the Overhauser field $B_{\rm N}(j)$.
We choose the Coulomb potential strength as position dependent:
\begin{equation} \label{eq:}
U(j) = U_0 \exp \left( - \frac{ (j/N)^6}{1- (j/N)^2} \right),
\end{equation}
again to smoothly interpolate between its full strength $U_0$ at the QPC center and the interaction-free leads. Finally, the 1D chain is attached to the semi-infinite leads in the local equilibrium with the Fermi distribution
$n_l(\epsilon)$ ($l={\rm S, D}$), given in Eq.~\eqref{eq:n}.

\begin{figure}
\includegraphics[width=\columnwidth]{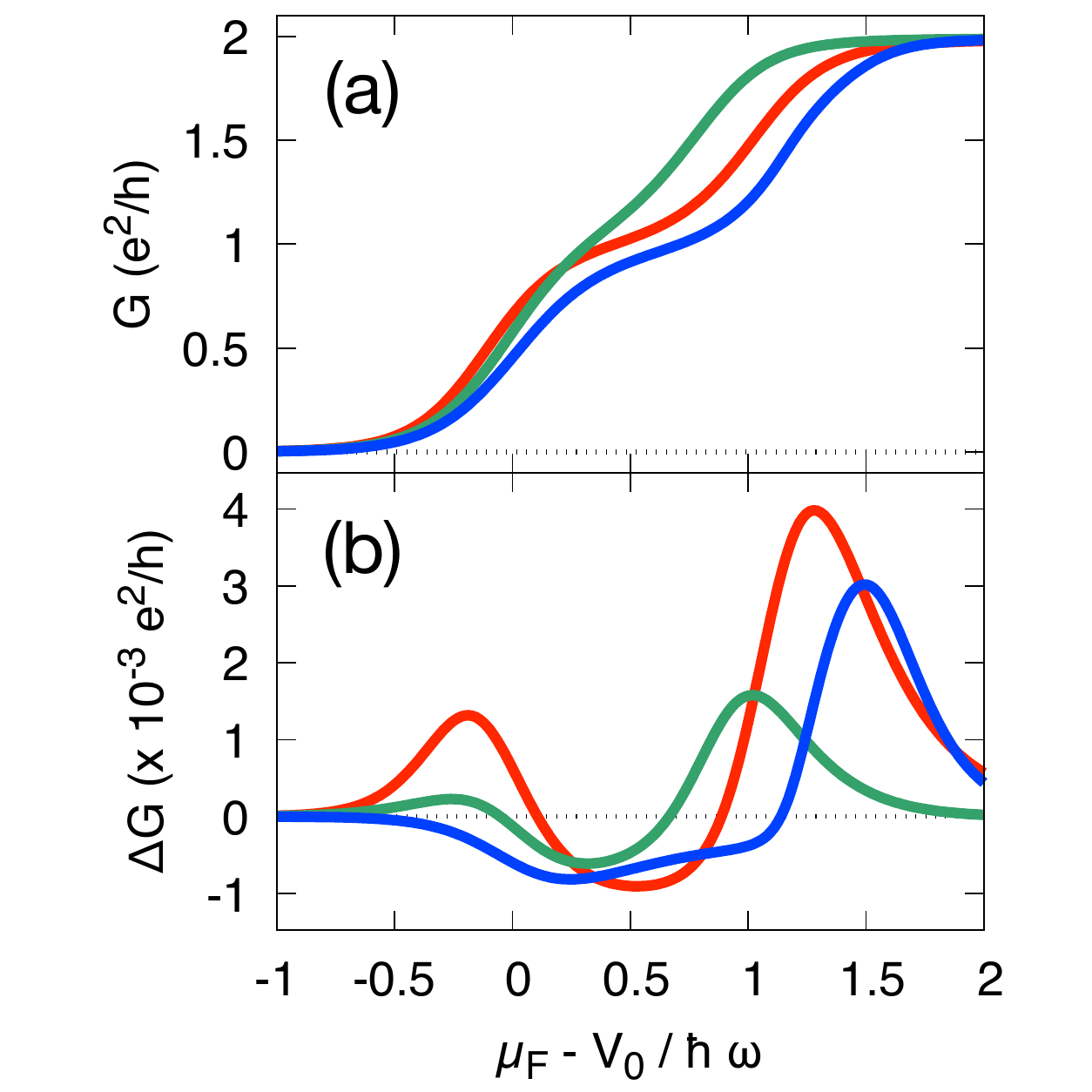}
\caption{\label{fig:dg-neg-other}
(a) Conductance and (b) its change upon partial depolarization of a dipole-like nuclear polarization versus $V_0$ for different electron-electron interactions strengths. The red/green/blue corresponds to $(U,\hbar \omega)=(7.7,0.5)/(10.2,1.0)/(12.8,1.0)$ meV. The other parameters are the same as in Fig.~\ref{fig:large_U}. 
}
\end{figure}

The interaction term is treated by a mean-field approximation neglecting spin fluctuations.
The mean-field spin density $\langle n_{j,\sigma} \rangle$ is  determined by
a self-consistent Green's function method \cite{Datta:2005:Book},
as follows. First, we evaluate $\langle n_{j,\sigma} \rangle$ for a given on-site energy $\epsilon_{\sigma}(j)$ using the correlation functions described in Chap.~9 of Ref.~\onlinecite{Datta:2005:Book}; at zero bias voltage, $\langle n_{j,\sigma} \rangle$  is evaluated using the local spectral function and the equilibrium Fermi-Dirac function while at finite bias voltages $\langle n_{j,\sigma} \rangle$  is evaluated via the partial spectral functions originating from the left and right leads with the Fermi-Dirac functions Eq.~\eqref{eq:n}.  
Then, $\epsilon_{\sigma}(j)$ is shifted by $U_j \langle n_{j,\bar{\sigma}} \rangle$
with $\bar{\sigma}$ the spin opposite to $\sigma$.
We then repeat the calculation on $\langle n_{j,\sigma} \rangle$ until convergence.
We calculate the magnetization density profile
$m_j =  \langle n_{j,\uparrow}  - n_{j,\downarrow} \rangle $ and the conductance/current through the QPC using Eq.~\eqref{eq:current}, the Landauer formula.

\begin{figure}
\begin{center}
\includegraphics[width=\columnwidth]{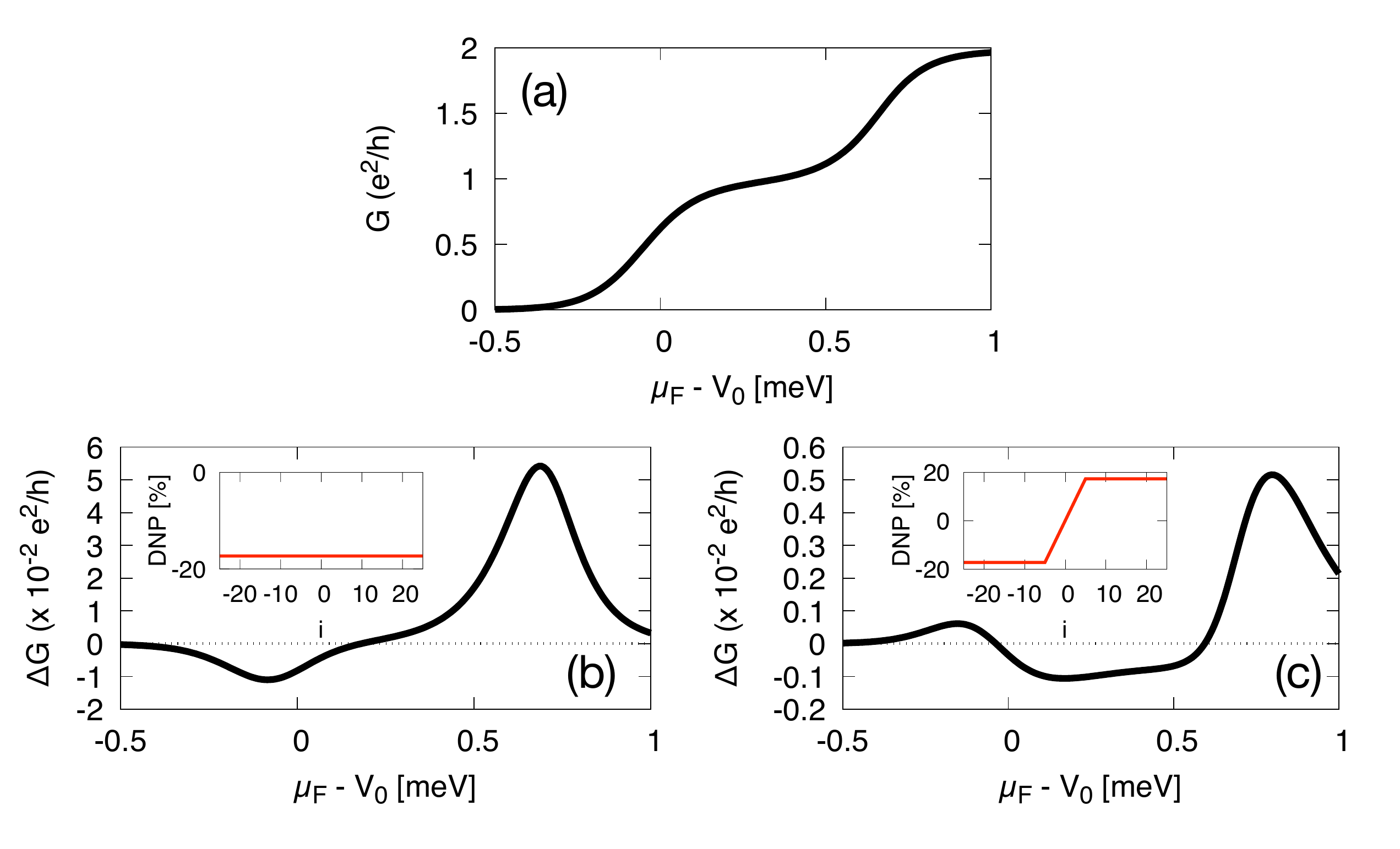}
\end{center}
\caption{\label{fig:large_U_uni}
Signal sign reversal for strong Coulomb interaction and dipole-like nuclear polarization. This figure is an analog to Fig.~\ref{fig:small_U} and we use the same parameters as there except for $U=7.7$ meV. (a) The conductance $G$.
(b), (c) $\Delta G$ calculated according to Eq.~\eqref{eq:DG} for $p=0.5$ and a (b) uniform and (c) dipole-like initial nuclear polarizations (shown in the insets). Compared to weaker Coulomb interactions [Figs.~\ref{fig:small_U}(b) and \ref{fig:small_U}(d)], a stronger interaction inverts the signal sign at the $G=1/2$ plateau only in panel (c), but not in panel (b).}
\end{figure}

\section{$\Delta G < 0$} 

\label{app:DGneg}

Here, we demonstrate that the sign reversal of the RD-NMR signal by strong interactions is robust. To this end, we repeat the calculation presented in Fig.~\ref{fig:large_U} for different values of the electron-electron interaction strength and the QPC potential curvature. Figure \ref{fig:dg-neg-other} shows the results. One can see that the region of the negative signal $\Delta G<0$ is correlated with the 1/2 conductance plateau, and therefore becomes more pronounced as the interaction strengths increases. 
Note that the strong Coulomb interaction does not induce qualitative changes for a uniform nuclear polarization as shown in Fig.~\ref{fig:large_U_uni}.

\section{Derivation of DNSP rates}

\label{app:rates}

Here, we derive Eq.~\eqref{eq:Iz1}. The derivation is a standard Fermi's golden rule calculation [see Chap. 5.3 in Ref.~\onlinecite{slichter} and Eq.~(1) in Ref.~\onlinecite{berg1990:PRL}]. We nevertheless find it useful to provide it here, as it is necessary to adapt these standard results for our system in which the electronic states are both spatially and spin dependent. To this end, let us start with the Fermi's golden rule formula, 
\be
W_{fi}^{(n)} = \frac{2\pi}{\hbar}  |\langle \{\Psi_f\} I_f | H_I^{(n)} | \{\Psi_i\} I_i \rangle |^2 \delta (E_i-E_f),
\label{eq:FGR}
\ee
for the rate of transition concerning the nuclear spin $n$, between the initial state $i$  and the final state $f$. Both of these many-particle states are composed of the electronic subsystem (described by the set of occupied scattering states $\{\Psi\}$) and the state $I$ of the nuclear spin $n$. We have accordingly specified to the part of the electron-nuclear interaction which pertains to nucleus $n$ only:
\be
H_I^{(n)}=  \frac{A v_0}{S_\perp} t_n \delta(x - x_n) \boldsymbol{\sigma} \cdot {\bf I}_n.
\ee
From here on we omit the nuclear spin index $n$. We rewrite the spin operator product in the previous as  
\be
\boldsymbol{\sigma} \cdot {\bf I} = \sigma_z I_z + \frac{1}{2}\sigma_+ I_- +\frac{1}{2}\sigma_-I_+,
\ee
where we have introduced $\sigma_\pm = \sigma_x \pm i \sigma_y$ and analogously for the nuclear spin operators. We notice that the three terms correspond to transitions which, respectively, do not change $I$, decrease it by 1, and increase it by 1. We are interested in the changes of the nuclear spin probabilities, to which only the latter two terms contribute. Summing over all initial states, occurring with probability $p_i$, and all available final states $f$ (except of the state $f=i$), the nuclear spin evolution follows as \be
\partial_t \langle I \rangle = W_{\rm inc} - W_{\rm dec},
\label{eq:step4}
\ee
the difference between rates increasing and decreasing the nuclear spin. If we now consider independent nuclear and electronic subsystems, so that $p_i=p^n_i p^e_i$, we can split the rates into their electronic and nuclear constituents
\begin{subequations}
\begin{eqnarray}
W_{\rm inc} &=& W_{\downarrow\uparrow} \sum_{i,f} p^n_i | \langle I_f | I_+ | I_i \rangle |^2,\label{eq:step5a}\\
W_{\rm dec} &=& W_{\uparrow\downarrow} \sum_{i,f} p^n_i | \langle I_f | I_- | I_i \rangle |^2,\label{eq:step5b}
\end{eqnarray}
\label{eq:step5}
\end{subequations}
where the electronic rates are
\be
\begin{split}
W_{\downarrow\uparrow} &= \frac{2\pi}{\hbar}\frac{A^2 v_0^2 t_n^2}{4S_\perp^2}\sum_{i,f} p^e_i  \times \\
\times &| \langle \{\Psi_f\} | \sigma_- \delta(x-x_n) | \{\Psi_i\} \rangle |^2 \delta(\epsilon_i- \epsilon_f+\epsilon^N_z).
\end{split} 
\label{eq:step6}
\ee
Here, we used that the nuclear subsystem energy change for this transition is always the same, equal to the nuclear Zeeman energy and the energies $\epsilon$ are energies of the electronic subsystem. The rate $W_{\uparrow\downarrow}$ is given by the same equation upon replacements $\sigma_- \to \sigma_+$ and $\epsilon^N_z \to -\epsilon^N_z$. In the many-particle matrix element, only such pairs of single-particle states contribute in which the initial state is occupied and the final state is empty, for which the probability $p^e $ results in the fermionic occupation factors
\be
\begin{split}
W_{\downarrow\uparrow} &= \frac{2\pi}{\hbar}\frac{A^2 v_0^2 t_n^2}{S_\perp^2}\sum_{a,a^\prime} |\Psi_{a^\prime\downarrow} (x_n) \Psi_{a\uparrow} (x_n)|^2 \times \\
&\times 
n_{a\uparrow} [1-n_{a^\prime\downarrow} ]
\delta(\epsilon_{a\uparrow}- \epsilon_{a^\prime\downarrow}+\epsilon^N_z).
\end{split} 
\label{eq:step7}
\ee
Here, we denote $a$ as all quantum numbers of the state except of the spin, which has been fixed by the $\sigma_-$ operator in Eq.~\eqref{eq:step6}. For the scattering states, according to Eq.~\eqref{eq:ss} these indices are the lead of origin and energy. The summation over the states is then explicitly 
\be
\sum_{a} = \sum_{l} \int {\rm d}\epsilon \, g_{l\sigma}(\epsilon),
\ee 
with the density of electronic states incoming toward the scatterer from lead $l$ with spin $\sigma$  at energy $\epsilon$ being
\be
g_{\l\sigma}(\epsilon) = \frac{1}{2\pi} \frac{1}{\hbar v_{l\sigma}(\epsilon)}.
\ee
We used a unit normalization volume, in accordance with Eq.~\eqref{eq:ss}, and the velocity defined in Eq.~\eqref{eq:kv}. Using it in Eq.~\eqref{eq:step7} gives
\be
\begin{split}
W_{\downarrow\uparrow} &= \frac{2\pi}{\hbar}\frac{A^2 v_0^2 t_n^2}{S_\perp^2} \sum_{l,l^\prime} \int {\rm d} \epsilon \, 
g_{l\uparrow}(\epsilon) g_{l^\prime\downarrow}(\epsilon^\prime)\times \\
&\times 
|\Psi_{\epsilon^\prime l^\prime \downarrow} (x_n) \Psi_{\epsilon l \uparrow} (x_n)|^2 
n_{l\uparrow}(\epsilon) [1-n_{l^\prime\downarrow}(\epsilon^\prime)],
\end{split} 
\label{eq:step9}
\ee
where the delta function gave $\epsilon^\prime = \epsilon+\epsilon^N_z$, which is Eq.~\eqref{eq:Wss}. 

We now move to the evaluation of the nuclear spin-related factors in Eqs.~\eqref{eq:step5a} and \eqref{eq:step5b}, which we denote as $p_+$ and $p_-$, respectively. Identifying $\sum_i | I_i\rangle p^n_i \langle I_i|$ with the density matrix $\rho_n$ of the nuclear spin $n$, these expressions are
\be
p_\pm = {\rm tr} \left( I_\pm \rho_n I_\mp \right).
\ee
Their difference is
\be
p_+-p_- = {\rm tr} \left( [I_-, I_+] \rho_n  \right) = -2 {\rm tr} \left( I_z \rho_n  \right) \equiv -2\Iz,
\label{eq:step11}
\ee
while the sum is
\be
p_+ + p_- = {\rm tr} \left( \{ I_- , I_+\} \rho_n  \right) = 2 {\rm tr} \left( {\bf I}^2 \rho_n - I_z^2 \rho_n  \right).
\ee
Here, we make the assumption that the polarization is small, so that the statistical average of the $I_z^2$ is equal to one third of the average of 
${\bf I}^2=I(I+1)$.\footnote{The replacement is an exact identify for $I=1/2$; alternatively, one could assume a thermal distribution for the nuclear spin---the nuclear spin temperature assumption---and calculate the expression analytically.} We get
\be
p_+ + p_- =  \frac{4}{3} I(I+1).
\label{eq:step13}
\ee 
Putting together Eqs.~\eqref{eq:step4}, \eqref{eq:step9}, \eqref{eq:step11}, and \eqref{eq:step13} gives Eq.~\eqref{eq:Iz1}.

\section{The microscopic equilibrium relations}

Here, we derive Eq.~\eqref{eq:eq}, the microscopic equilibrium relation for the nuclear relaxation rate, and show that an analogous equation for the pumping rate is different, in general. To this end, we start with Eq.~\eqref{eq:W1} in which, for generality, we keep the possible lead dependence of the density of states and non-zero spin voltages in the leads:
\be
\begin{split}
W_{\overline{\sigma}\sigma}^{(1)} & = W_0  t_n^2 \sum_{l} \int \frac{{\rm d}\epsilon}{\hbar} \frac{v_F^2}{v_{l \sigma}(\epsilon)  v_{l \overline{\sigma}}(\epsilon^\prime)}
\\&\times|\Psi_{\epsilon l \sigma}(x_n)|^2 |\Psi_{\epsilon^\prime l \overline{\sigma}}(x_n)|^2 n_{l\sigma}(\epsilon) [1 - n_{l\overline{\sigma}}(\epsilon^\prime) ].
\end{split}
\label{eq:W1_gen1}
\ee
Note that here $\epsilon^\prime=\epsilon-\sigma\epsilon_z^N$. Using Eq.~\eqref{eq:nsym} we get
\be
\begin{split}
W_{\overline{\sigma}\sigma}^{(1)} & = W_0  t_n^2 \sum_{l} \int \frac{{\rm d}\epsilon}{\hbar} \frac{v_F^2}{v_{l \sigma}(\epsilon^{\prime \prime}) v_{l \overline{\sigma}}(\epsilon)}
\\&\times |\Psi_{\epsilon^{\prime \prime} l \sigma}(x_n)|^2 |\Psi_{\epsilon l \overline{\sigma}}(x_n)|^2 n_{l\overline{\sigma}}(\epsilon) [1 - n_{l\sigma}(\epsilon^{\prime \prime}) ]\\
&\times \exp\left[-\sigma \beta (\epsilon_z^N -2 \delta \mu_l)\right],
\end{split}
\label{eq:W1_gen2}
\ee
where we denoted $\epsilon^{\prime \prime} = \epsilon+\sigma \epsilon_z^N$. Except for the term in the last line, the expression is equal to $W^{(1)}_{\sigma\overline{\sigma}}$, the relaxation rate for flipped spin indexes. If there are no spin voltages, $\delta \mu_l=0$, we obtain Eq.~\eqref{eq:eq}. For finite spin voltages, we get from Eq.~\eqref{eq:W1_gen2} that the relaxation rate is biased, as the electron-nuclear equilibration is towards a distribution with an effective nuclear Zeeman energy to which the electron spin voltages contribute. This is in line with the results of Sec.~\ref{sec:non-eq}. Note, however, that there we use slightly different meaning for the upper indices on the rates. Namely, the spin-voltage part of Eq.~\eqref{eq:W1_gen2} is assigned to the rate ``II'' in Sec.~\ref{sec:non-eq}. 

For illustration, we now perform the same steps for the pumping rate $W^{(2)}$ [Eq.~\eqref{eq:W2}]. Starting with
\be
\begin{split}
W_{\overline{\sigma}\sigma}^{(2)} & = W_0  t_n^2 \sum_{l} \int \frac{{\rm d}\epsilon}{\hbar} \frac{v_F^2}{v_{l \sigma}(\epsilon) v_{\overline{l} \overline{\sigma}}(\epsilon^\prime)}
\\&\times |\Psi_{\epsilon l \sigma}(x_n)|^2 |\Psi_{\epsilon^\prime \overline{l} \overline{\sigma}}(x_n)|^2 n_{l\sigma}(\epsilon) [1 - n_{\overline{l}\overline{\sigma}}(\epsilon^\prime) ].
\end{split}
\label{eq:W2_gen1}
\ee
we transform it into
\be
\begin{split}
W_{\overline{\sigma}\sigma}^{(2)} & = W_0  t_n^2 \sum_{l} \int \frac{{\rm d}\epsilon}{\hbar} \frac{v_F^2}{v_{l \sigma}(\epsilon^{\prime \prime}) v_{\overline{l} \overline{\sigma}}(\epsilon)}
\\&\times |\Psi_{\epsilon^{\prime \prime} l \sigma}(x_n)|^2 |\Psi_{\epsilon \overline{l} \overline{\sigma}}(x_n)|^2 n_{\overline{l}\overline{\sigma}}(\epsilon) [1 - n_{l\sigma}(\epsilon^{\prime \prime}) ]\\
&\times\exp\left[-\sigma \beta (\epsilon_z^N - \delta \mu_{\overline{l}}+\delta \mu_{l})+\beta (\mu_{l}-\mu_{\overline{l}})\right].
\end{split}
\label{eq:W2_gen2}
\ee
For zero spin voltages, the relation in Eq.~\eqref{eq:W2_gen2} contains, compared to the one in Eq.~\eqref{eq:W1_gen2}, additional factors 
$\exp(\pm \beta eV)$. If a bias voltage is applied, it typically dominates the nuclear Zeeman energies. On the other hand, for zero bias and zero spin voltages, the rate $W^{(2)}$ also fulfills Eq.~\eqref{eq:eq}, as we mentioned below Eq.~\eqref{eq:eq}.

\section{Net nuclear polarization due to the leads velocities or crosssections asymmetry}

\label{app:geometry asymmetry}

We now consider possible asymmetries in the source and drain leads. We are interested in the degree of violation of the symmetry relations in Eqs.~\eqref{eq:w2sym} and \eqref{eq:main1}. Let us first consider different velocities in the two leads. At a fixed total energy, such a difference arises if the potential bottom of the two leads is not the same. The applied voltage naturally leads to such a potential drop. If the total charge density in a lead is fixed, the difference of the potential bottoms is the same as the applied voltage. The difference of the velocities in the two leads is then of the order of $eV/\mu_{\rm F}$. For a typical Fermi energy of 10 meV and bias voltage 100 $\mu$eV, the velocities differ by roughly a factor of 10$^{-2}$ on the relative scale.

In this case, Eq.~\eqref{eq:Psisym} should be replaced by
\be
v_{l\sigma} |\Psi_{\epsilon l \sigma}(x)|^2 = v_{\overline{l}\sigma} |\Psi_{\epsilon \overline{l} \sigma}(-x)|^2,
\label{eq:Psisym_gen}
\ee
stating that the scattering matrix unitarity gives relations for the current densities, rather than the particle densities. Using this relation in Eq.~\eqref{eq:W2}, we would get additional factors 
\be
\frac{v_{\overline{l}\sigma}v_{l\overline{\sigma}}}{v_{l\sigma}v_{\overline{l}\overline{\sigma}}} \approx 1 \pm 2\frac{\epsilon_z eV}{\mu_{\rm F}^2},
\ee
arising in the equation analogous to Eq.~\eqref{eq:w2sym}. These factors differ from one by a negligibly small amount, perhaps 10$^{-3}$ or 10$^{-4}$, and therefore the velocity effects can be safely neglected.

We now look at the effects of lead geometrical asymmetry, meaning that the source and drain leads are different. We estimate such effects roughly by considering the consequences upon scaling the cross section profile along one of the transverse directions (having in mind a 2DEG, the growth direction profile is fixed, while the lateral transverse one might differ) by a dimensionless factor $\xi$:
\be
\rho(x,y,z) \to \rho^\prime(x,y,z) = \xi^{-1} \rho(x,y/\xi,z).
\ee
With the chosen prefactor, the new profile is correctly normalized
\be
\int {\rm d}y {\rm d}z \rho^\prime(x,y,z) = 1.
\ee
We find that the electron-related DNSP rates [Eq.~\eqref{eq:W1W2}] are multiplied by $\xi^{-2}$, due to the change in the cross section $ S_\perp^\prime = S_\perp \xi$. Noting that the factors $t_n$ [Eq.~\eqref{eq:tn}] and $A^\prime$ [Eq.~\eqref{eq:Ap}] are not changed by the coordinate rescaling, we get the following: First, if the nuclear relaxation rate is dominated by electrons, so that the rates $W^{(0)}$ can be neglected, the rescaling does not affect the nuclear polarization $\Iz$ [Eq.~\eqref{eq:Iz3}] nor the Overhauser energy $\delta \epsilon_z$ [Eq.~\eqref{eq:dez}]. If, on the other hand, the relaxation is dominated by other channels than the electrons, both the nuclear polarization, and the Overhauser energy are multiplied by factor $\xi^{-2}$. As an example, if the source lead is twice wide compared to the drain lead, an approximately four times smaller polarization and Overhauser energy follows. We conclude that the geometrical-related asymmetry is expected to dominate the velocity (in fact, the applied bias) asymmetry.

\end{document}